\documentclass[12pt]{article}
\usepackage{fullpage}
\usepackage{cite}
\usepackage{amstext,amsfonts,amsbsy,eucal,amssymb}
\usepackage{epsfig}
\usepackage{psfrag}

\newcounter{definition}

\def\thefigure{\arabic{section}.\arabic{figure}}
\def\theequation{\thesection.\arabic{equation}}
\def\appendix{
  \setcounter{section}{0}
  \setcounter{subsection}{0}
  \par
  \def\thesection{Appendix \Alph{section}}
  \def\theequation{\Alph{section}.\arabic{equation}}
  \def\thefigure{\Alph{section}.\arabic{figure}}
}
\makeatletter
\def\fnum@figure{Fig. \thefigure}
\@addtoreset{equation}{section}
\@addtoreset{figure}{section}
\makeatother

\title{Partition Function \\for a 1-D $\delta$-function Bose Gas} 

\author{Go {\sc Kato}\thanks{go-karto@monet.phys.s.u-tokyo.ac.jp} \, and Miki
  {\sc Wadati}\thanks{wadati@monet.phys.s.u-tokyo.ac.jp} \\ Department of Physics,
  Graduate School of Science, \\ University of Tokyo, Hongo 7-3-1,
  Bunkyo-ku, Tokyo 113-0033 }
\date{May 26, 2000}

\begin{document}
\maketitle
\setlength{\baselineskip}{1.8em}

\begin{abstract}
\setlength{\baselineskip}{2em} The $N$-particle partition function of
a one-dimensional $\delta$-function bose gas is calculated explicitly
using only the periodic boundary condition (the Bethe ansatz
equation). The $N$-particles cluster integrals are shown to be the
same as those by the thermal Bethe ansatz method. \\ \\ keywords:
thermal Bethe ansatz method, $\delta$-function bose gas, integrable
system, partition function, cluster expansion
\end{abstract}

\newpage

\section{Introduction}
Extending the work by Lieb and Liniger~\cite{lieb}, Yang and
Yang~\cite{yang} presented an ingenious method to study the equilibrium
thermodynamics of one-dimensional system of bosons with repulsive
$\delta$-function interaction. The Hamiltonian of the system is
\begin{eqnarray}
H_N&=&-\sum_{i=1}^{N}
       \left(
          \frac{\partial^2}{\partial x_i^2}
         +\kappa \sum_{j \neq i}\delta \left(x_j-x_i\right)
       \right),
\label{eq:Hamiltonian}
\end{eqnarray}
where $\kappa$ is the coupling constant and is assumed to be positive.
Throughout the paper, the Planck constant and the mass of a particle
are chosen to be $\hbar = 1$ and $2m = 1$.  This system is a first
quantized version of the nonlinear Schr\"odinger model described by
the Hamiltonian operator,
\begin{eqnarray}
H=\int dx
  \left[
    \phi_x^+\phi_x+\kappa\phi^+\phi^+\phi\phi
  \right],
\end{eqnarray}
where $\phi\left(x,t\right)$ and $\phi^+\left(x,t\right)$ are  boson field operators. The method
invented by Yang and Yang is called the thermal Bethe ansatz (TBA)
method since it starts from the Bethe ansatz wave function
\cite{bethe} and enables us to calculate the thermodynamic quantities
at finite temperature. While there have been many successful
applications to quantum particle and spin systems, not much study on
the method itself has been done. A crucial assumption is the form of
the entropy. Thacker\cite{thacker} studied the $\delta$-function bose
gas in infinite volume by the field-theoretic perturbation method and
reproduced the results of the TBA method. One of the authors
(M.W.)~\cite{wadati_b} presented a bosonic formulation of the TBA
method and calculated the grand partition function of the system at
infinite volume. In this paper, we calculate explicitly the
$N$-particle partition function of the
Hamiltonian~(\ref{eq:Hamiltonian}) under the periodic boundary
condition.  That is, the assumptions of the TBA method are not used
and all the calculations are carried out exactly at finite
volume. A preliminary result for $N=2,3,4$ has been reported in
\cite{go_1}.

The outline of the paper is the following. In \S2, we present a new
method to evaluate the partition function for the $N$-particle system. A key idea is that we use only the periodic
boundary condition of the wave function.
In \S3, we calculate the $N$-particle cluster integrals using the partition function.
An explicit calculation for the $N$-particle
system has been done for the first time. 
The results are compared
 with those given by the TBA. We prove that both results
completely agree.
The last section is devoted to concluding remarks.
Technical details of calculations are summarized in \ref{sec:Proof_on_summation}$\sim$D.

\section{$N$-particle partition function}
We study a statistical mechanics of the quantum $N$-particle
system~(\ref{eq:Hamiltonian}).
Let $E$, $L$ and $\left\{k_i\right\}$ denote the total energy, the
system size and the wave numbers. It is known that the total energy and the wave numbers are
determined by the following relations,
\begin{eqnarray}
E    &  =& \sum_{i=1}^{N}k_i^2,
\label{eq:bethe_1}\\
Lk_i &  =& 2{\pi}n_i
          +\sum_{j<i}\Delta\left(k_j-k_i\right)
          -\sum_{j>i}\Delta\left(k_i-k_j\right),
\label{eq:bethe_2}
\label{eq:boundary_condition}
\end{eqnarray}
where $\left\{n_i\right\}$ are integers satisfying the condition,
\begin{eqnarray}
n_i &\ge& n_{i+1},
\label{eq:condition_of_n}
\end{eqnarray}
and
$\Delta\left(k\right)$ is the phase-shift for two-body scattering,
\begin{eqnarray}
\Delta\left(k\right)&  = & -2\tan^{-1}\left(\frac{\kappa}k\right).
\label{eq:bethe_3}
\label{eq:define_delta}
\end{eqnarray}
The relation (\ref{eq:bethe_2}) is obtained from the periodic boundary
condition imposed on the $N$-particle eigenfunction and is called the
Bethe ansatz equation. The range of the
function~(\ref{eq:define_delta}) is assumed to be $\left(-2\pi ,
  0\right)$, and this phase-shift is not a ``true'' one
in the sense of \cite{wadati_b}. That is, the phase-shift
function~(\ref{eq:define_delta}) has analyticity on the real axis.
The appearance of only two-body S-matrix in (\ref{eq:boundary_condition}) implies the factorization of $S$-matrices, which is one of the remarkable properties of integrable systems.

The partition function for the $N$-particle system is defined by
\begin{eqnarray}
 Z_N 
& = & 
 \sum_{\left\{n_j\right\}} \exp\left(-\beta \sum_{i=1}^{N}k_i^2\right),
\label{eq:N_partition_function}
\end{eqnarray}
where, with $T$ being the absolute temperature,
\begin{eqnarray}
\beta = \frac{1}{k_BT},
\end{eqnarray}
and the summation is over all possible configurations of $\left\{n_i
\right\}$ under the condition (\ref{eq:condition_of_n}). Without recourse to the
TBA method, we calculate the partition function $Z_N$ only by
using the relations (\ref{eq:bethe_1})$\sim$(\ref{eq:bethe_3}). We explain a method in three steps.
\subsection{Change of summation}
In terms of new variable and function,
\begin{eqnarray}
 {\tilde{n}}_m & \equiv & n_m - m + \frac{N+1}{2},
\label{eq:define_n_and_n_tilde}  \\
 {\tilde{\Delta}\left(k\right)} & \equiv & \Delta\left(k\right)+\pi,
\end{eqnarray}
the periodic boundary condition (\ref{eq:bethe_2}) is changed into
\begin{eqnarray}
Lk_i &  =& 2{\pi}{\tilde n}_i
          +\sum_{j \neq i}{\tilde \Delta}\left(k_j-k_i\right).
\label{eq:bethe_2_m1}
\end{eqnarray}
We interpret (\ref{eq:bethe_2_m1}) as analytic relations
between real numbers $\left\{\tilde{n}_i\right\}$ and $\left\{k_i\right\}$.
From the symmetry of (\ref{eq:bethe_2_m1}), we see that, when 
$\left\{\tilde{n}_1,\cdots,\tilde{n}_i,\cdots,\tilde{n}_j,\cdots,\tilde{n}_N \right\}$
corresponds to $\left\{k_1,\cdots,k_i,\cdots,\right.$\\
$\left.k_j,\cdots,k_N\right\}$,
$\left\{\tilde{n}_1,\cdots,\tilde{n}_j,\cdots,\tilde{n}_i,\cdots,\tilde{n}_N \right\}$
 should correspond to
$\left\{k_1,\cdots,k_j,\cdots,k_i,\cdots,k_N\right\}$.

\begin{figure}[p]
\begin{center}
\begin{psfrags}
  \psfig{file=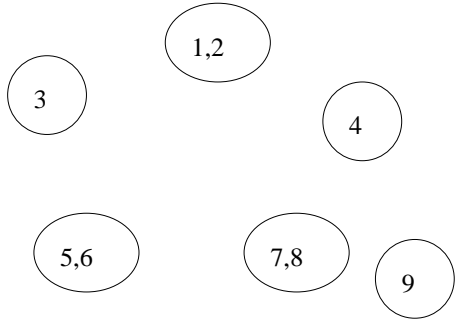 , scale = 0.55}
\end{psfrags}
\end{center}
\caption{A graphical representation of a set $\theta_{9}$ in (\ref{eq:example_theta}).}
\label{fig:theta}
\end{figure}

It is convenient to introduce the following set function,
\begin{eqnarray}
\Theta \left( \sigma \right)
&\equiv&
\left\{ 
  \theta
  \left|
    \begin{array}{c}
      n\in \sigma,\quad
      \sigma^{\prime} \in \theta,\quad
      n \in \sigma^{\prime}\\
      \sigma'' , \sigma''' \in \theta, \quad
      \sigma''\neq \sigma''' \quad
      \Rightarrow \quad
      n'' \in \sigma'', \quad
      n'''\in \sigma''',\quad
      n'' \neq n'''
    \end{array}
  \right.
\right\}
.
\end{eqnarray}
A domain $\sigma$ of the function is a family of arbitrary finite
sets. The first condition, $n\in\sigma$, $\sigma'\in\theta$,
$n\in\sigma'$, means that the sum of the elements in $\theta$ is
$\sigma$. The second condition, $\sigma'',\sigma'''\in\theta$,
$\sigma''\neq\sigma'''$ $\Rightarrow$ $n''\in\sigma''$, $n'''\in
\sigma'''$, $n''\neq n'''$, means that the sum is a direct sum. That
is, the image $\Theta\left(\sigma\right)$ is all the families of sets
whose direct sum is the set $\sigma$.
If the argument is integer, we
define
\begin{eqnarray}
\Theta \left( N \right)
&\equiv&
\Theta \left( \left\{1,2,\cdots,N\right\}\right).
\label{eq:define_Theta}
\end{eqnarray}
For example,
\begin{eqnarray}
\Theta\left(9\right)
&\ni&
\left\{
\left\{1,2\right\},
\left\{3\right\},
\left\{4\right\},
\left\{5,6\right\},
\left\{7,8\right\},
\left\{9\right\}
\right\}=\theta_{9}.
\label{eq:example_theta}
\end{eqnarray}
Figure \ref{fig:theta} illustrates $\theta_9$ in (\ref{eq:example_theta}).
And we define a symbol, for arbitrary function
$f\left(n_1,\cdots,n_N\right)$,
\begin{eqnarray}
\sum\left(\theta,\left\{n_i\right\}\right)
f\left(n_1,\cdots,n_N\right)
&\equiv&
\sum_{\left\{n_{\sigma}\right\},\sigma\in\theta}
f\left(n_1,\cdots,n_N\right),\\
\theta\in\Theta\left(N\right),n_i=n_{\sigma},i\in\sigma.\nonumber
\label{eq:define_Sigma}
\end{eqnarray}
This symbol,
$\sum\left(\theta,\left\{n_i\right\}\right)f\left(n_i\right)$,
indicates that a function $f$ is summed up over all $n_i$ on the
condition that $n_j = n_k$ if an element which contains both $j$ and
$k$ exists.

The following formula can be proved (\ref{sec:Proof_on_summation}),
\begin{eqnarray}
\label{eq:sumrule_h}
\sum_{n_1< \cdots < n_N} f\left(n_1,\cdots,n_N\right)
&=&
\frac{1}{N!}
\sum_{\theta \in \Theta\left(N\right)}
F\left(\theta\right)
\sum\left(\theta,\left\{n_i\right\}\right)
f\left(n_1,\cdots,n_N\right)
\end{eqnarray}
where
\begin{eqnarray}
F\left(\theta\right)
&=&
\prod_{\sigma\in \theta}
\left(-1\right)^{\left(\sharp \sigma -1\right)}
\left(\sharp \sigma -1\right)!,
\label{eq:define_F}
\end{eqnarray}
\begin{eqnarray}
f\left(n_1,\cdots,n_i,\cdots,n_j,\cdots,n_N\right)
&=&
f\left(n_1,\cdots,n_j,\cdots,n_i,\cdots,n_N\right).
\end{eqnarray}
Here and hereafter, the number of elements in a set $\sigma$ is
denoted by $\#\sigma$.

By use of the  formula(\ref{eq:sumrule_h}), the partition function is
rewritten as
\begin{eqnarray}
Z_N 
& = &
\sum_{\cdots>\tilde{n}_i > \tilde{n}_{i+1}>\cdots} 
\exp\left(-\beta\sum_{j=1}^{N}k_j^2\right)  \nonumber\\
& = &
\frac{1}{N!}
\sum_{\theta \in \Theta\left(N\right)}
F\left(\theta\right)
\sum\left(\theta,\left\{\tilde{n}_i\right\}\right)
\exp\left(-\beta\sum_{j=1}^{N}k_j^2\right).
\label{eq:partition_function_m1}
\end{eqnarray}
\subsection{Replacement of summations by integrals}
To evaluate (\ref{eq:partition_function_m1}), we have a useful formula,
\begin{eqnarray}
\sum_{n = {\rm integer} }f\left(n\right)
& = & 
\sum_{n = {\rm integer}} \int_{-\infty}^{\infty} dn^{\prime} 
f\left(n^{\prime}\right)
\exp\left(-2\pi i n n^{\prime}\right),           \nonumber\\
&&{\rm if} \; \left|f\left(n\right)\right|< \exp\left(-r n^2\right) 
, r > 0.
\label{eq:summation_formula}
\end{eqnarray}
This can be proved by the Fourier transform and the Jacobi's
imaginary transformation for elliptic theta functions.  Applying the
formula (\ref{eq:summation_formula}), we replace summations in
(\ref{eq:partition_function_m1}) by integrals,
\begin{eqnarray}
Z_N
&=&
\frac{1}{N!}
\sum_{\theta \in \Theta\left(N\right)}
F\left(\theta\right)
\sum_{\left\{n_{\sigma}\right\},\sigma\in\theta}
\left(-1\right)^{\left(N-1\right)\sum_{\sigma\in\theta}n_{\sigma}}
\int\prod_{\sigma\in\theta}d\tilde{n^{\prime}}_{\sigma}
\nonumber\\&&{}\times
\exp
\left(
  \sum_{\sigma\in\theta}
  \left(
    -\beta k_{\theta}^2
    -2\pi i n_{\sigma} \tilde{n^{\prime}}_{\sigma}
  \right)
\right).
\label{eq:partition_function_m2}
\end{eqnarray}
To proceed further, we define four symbols.

\begin{figure}[p]
\begin{center}
\begin{psfrags}
  \psfig{file=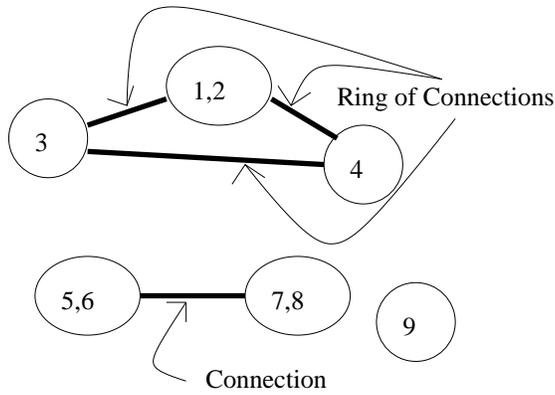 , scale = 0.55}
\end{psfrags}
\end{center}
\caption{The connection and the ring of connection.}
\label{fig:ring}
\end{figure}

\begin{figure}[p]
\begin{psfrags}
\begin{center}
  \psfig{file=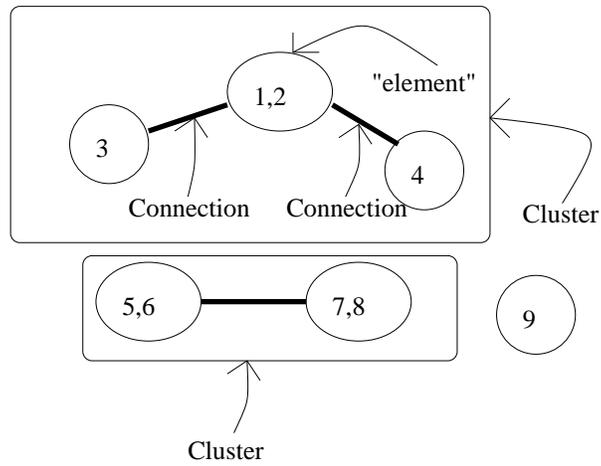 , scale = 0.55}
\end{center}
\end{psfrags}
\caption{The pattern of connection $\left[\lambda_9,\theta_{9}\right]$ in (\ref{eq:example_upsilon}). Here, $\left[\lambda_9,\theta_{9}\right]$ indicates that $\lambda_9$ is consistent with $\theta_{9}$}
\label{fig:pattern}
\end{figure}

First,
\begin{eqnarray}
\Lambda\left( \theta \right)
&\equiv&
\left\{
  \lambda
  \left|
    \begin{array}{c}
      \lambda = 
      \left\{
        \xi |
        \xi = \left\{\sigma , \sigma^{\prime} \right\}
        \quad \sigma , \sigma^{\prime} \in \theta 
      \right\}\\
      \lambda^{\prime} \subseteq \lambda
      \quad \sharp
      \left\{
        \sigma  |
        \xi \in \lambda^{\prime} 
        \quad\sigma\in \xi
      \right\}
      >
      \sharp\lambda^{\prime}
    \end{array}
  \right.
\right\},
\label{eq:define_Theta_tilde}
\\
&&\theta = \left\{\sigma_1,\sigma_2,\cdots\right\}.\nonumber
\end{eqnarray}
The set function $\Lambda\left( \theta \right)$ consists of elements
$\lambda$. $\lambda$ contains elements $\xi$, each of which has two
elements $\sigma,\sigma^\prime$. We call $\xi$ connection between the
two elements $\sigma,\sigma^\prime$. The connectivity is referred to
as pattern of connections. $\Lambda\left( \theta \right)$ represents a
set of all the patterns of connections that have no ring of
the connections. In Fig.\ref{fig:ring}, the connection and ring of
connections are illustrated.
For example of $\Lambda\left(\theta\right)$,
\begin{eqnarray}
\Lambda\left(\theta_{9} \right)
&\ni&
\left\{
\left\{
\left\{1,2\right\},
\left\{3\right\}
\right\}
,
\left\{
\left\{1,2\right\},
\left\{4\right\}
\right\}
,
\left\{
\left\{5,6\right\},
\left\{7,8\right\}
\right\}
\right\}
=
\lambda_{9}.
\label{eq:example_upsilon}
\end{eqnarray}
The pattern of connection $\lambda_9$ is shown in Fig. \ref{fig:pattern}.
As a special case, $\Lambda\left( \theta \right)$ contains a pattern of no
connection. i.e. $\Lambda\left( \theta \right) \ni \emptyset$.

Second,
\begin{eqnarray}
n\left( \sigma , \lambda \right)
&\equiv&
\sharp
\left\{\xi
  \left| 
    \xi = 
    \left\{
      \sigma,\sigma^{\prime}
    \right\},\quad
    \sigma^{\prime}\in \left\{\sigma_1,\sigma_2,\cdots \right\},\quad
    \xi \in \lambda
  \right.
\right\},
\label{eq:define_n}
\\
&&
\sigma \in \left\{\sigma_1,\sigma_2,\cdots\right\},\quad
\lambda \in 
\Lambda\left(
  \left\{\sigma_1,\sigma_2,\cdots\right\}
\right).\nonumber
\end{eqnarray}
In words,
$n\left( \sigma , \lambda \right)$ is a number of elements which are linked with $\sigma$.

Third,
\begin{eqnarray}
G\left(\lambda^{\prime}\right)
&\equiv&
\left\{
  \lambda \subseteq \lambda^{\prime}
  \left|
    \left\{\sigma^{\prime},\sigma^{\prime\prime}\right\} \in  \lambda,\quad
    \left\{\sigma^{\prime\prime},\sigma^{\prime\prime\prime}\right\} \in \lambda^{\prime}\quad
    \Rightarrow\quad
    \left\{\sigma^{\prime\prime},\sigma^{\prime\prime\prime}\right\} \in \lambda
  \right.
\right\},
\label{eq:define_G}
\\
&&
\lambda^{\prime} \in 
\Lambda\left(
  \left\{\sigma_1,\sigma_2,\cdots\right\}
\right).\nonumber
\end{eqnarray}
That is, $G\left(\lambda^{\prime}\right)$ means a set of connection
patterns each of which is a cluster of the connection pattern
$\lambda^{\prime}$. Here, the cluster means that if two
connections $\xi$ and $\xi^{\prime}$ in
$\lambda^{\prime}$ are linked with a common element~$\sigma$, then $\xi$ and $\xi^{\prime}$ are in
the same cluster.
For example,
\begin{eqnarray}
G\left(\lambda_{9} \right)
&\ni&
\left\{
\left\{
\left\{1,2\right\},
\left\{3\right\}
\right\}
,
\left\{
\left\{1,2\right\},
\left\{4\right\}
\right\}
\right\}.
\end{eqnarray}
 A pattern of no connection is included in $G\left(\lambda^{\prime}\right)$.

Fourth,
\begin{eqnarray}
g \left( \left[\lambda ,\theta\right] \right)
&\equiv&
\left\{
  \sigma
  \left|
    \begin{array}{c}
      \sigma\in\theta,\quad
      \lambda^{\prime}\in G\left(\lambda\right),\quad
      \xi\in  \lambda,\quad
      \sigma\not\in \xi
      \\
      {\rm or}\\
      \lambda^{\prime}\in G\left(\lambda\right),\quad
      \lambda^{\prime}\neq\emptyset,\quad
      \sigma =
      \bigcup_{\left\{\sigma^{\prime},\sigma^{\prime\prime}\right\} \in \lambda^{\prime}}
      \left(
        \sigma^{\prime}\cup\sigma^{\prime\prime}
      \right)
\end{array}
  \right.
\right\},
\label{eq:define_g}
\\
&&
\theta = \left\{\sigma_1,\sigma_2,\cdots\right\}
,
\lambda \in 
\Lambda\left(
  \theta
\right).\nonumber
\end{eqnarray}
$\left[ \lambda ,\theta\right]$ means that $\lambda$ is consistent
with the connection pattern of $\theta$; An element which is linked with $\lambda$ is in  $\theta$.
$g \left(\left[ \lambda ,
    \theta\right] \right)$ indicates a set of direct sum ``elements of
$\;\theta$'' which are connected by one cluster in the connection
pattern $\lambda$. The reason why an element is quoted here is that the
element, at the same time, is a set. If an element of $\theta$ has no
connection, the ``element'' belongs to $g \left(\left[ \lambda ,
    \theta\right] \right)$.  For example,
\begin{eqnarray}
g\left(\left[\lambda_{9},\theta_{9}\right]\right)
&=&
\left\{
\left\{1,2,3,4\right\},
\left\{5,6,7,8\right\},
\left\{9\right\}
\right\}.
\end{eqnarray}

The following formula is proved (see a proof in
\ref{sec:Proof_on_Jacobian}):
\begin{eqnarray}
\left|A_{\sigma,\sigma^{\prime}}\left(\theta\right)\right|
&=&
\sum_{\lambda\in \Lambda\left(\theta\right)}
\left(
  I\left(\left[\lambda,\theta\right]\right)
  \prod_{\left\{\sigma,\sigma^{\prime}\right\}\in \lambda}
  x_{\sigma,\sigma^{\prime}}
\right),
\label{eq:jacobian_h}
\end{eqnarray}
with
\begin{eqnarray}
I\left(\left[\lambda,\theta\right]\right)
&\equiv&
L^{N-\sharp \lambda}
\left(
  \prod_{\sigma\in \theta}
  \sharp \sigma^{n\left(\sigma,\lambda\right)-1}
\right)
\left(
  \prod_{\sigma^{\prime} \in g\left(\left[\lambda,\theta\right]\right)}
  \sharp \sigma^{\prime}
\right),
\label{eq:define_I}\\
A_{\sigma,\sigma^{\prime}}\left(\theta\right)
&=&
\left\{
  \begin{array}{ll}
    L+\sum_{\sigma^{\prime\prime} \in \theta}
    \sharp \sigma^{\prime\prime}
    \; x_{\sigma ,\sigma^{\prime\prime}}&{\rm if}\quad \sigma =\sigma^{\prime}\\
    -\sharp \sigma^{\prime}
    \;x_{\sigma,\sigma^{\prime}} &{\rm if}\quad \sigma \neq\sigma^{\prime},
  \end{array}
\right.
\label{eq:define_A}
\end{eqnarray}
\begin{eqnarray}
\theta \in \Theta\left(N\right),\quad
\sigma,\sigma'\in\theta,\quad
x_{\sigma,\sigma^{\prime}}\equiv x_{\sigma^{\prime},\sigma},\quad
x_{\sigma,\sigma}\equiv 0.\nonumber
\end{eqnarray}

  Using the formula(\ref{eq:jacobian_h}), we change variables in (\ref{eq:partition_function_m2}) from
$\left\{\tilde{n}_i\right\}$ to $\left\{k_i\right\}$. In terms of the above introduced symbols, we obtain
\begin{eqnarray}
Z_N
&=&
\frac{1}{N!}
\sum_{\theta \in \Theta\left(N\right)}
F\left(\theta\right)
\sum_{\left\{n_{\sigma}\right\},\sigma\in\theta}
\left(-1\right)^{\left(N-1\right)\sum_{\sigma\in\theta}n_{\sigma}}
\nonumber\\&&\times{}
\int\prod_{\sigma\in\theta}\frac{dk_{\sigma}}{2\pi}
\sum_{\lambda\in \Lambda\left(\theta\right)}
I\left(\right[\lambda,\theta\left]\right)
\left(
  \prod_{\left\{\sigma,\sigma^{\prime}\right\}\in \lambda}
  \frac{d\Delta}{dk}\left(k_{\sigma}-k_{\sigma^{\prime}}\right)
\right)
\nonumber\\&&\times{}
\exp
\left[
  \sum_{\sigma\in\theta}
  \left\{
    -\beta \#\sigma k_{\sigma}^2
    -i n_{\sigma}
    \left(
      Lk_{\sigma}
      -\sum_{\sigma^{\prime}\in\theta,\neq \sigma }
      \#\sigma^{\prime}\tilde{\Delta}\left(k_{\sigma^{\prime}}-k_{\sigma}\right)
    \right)
  \right\}
\right]. 
\label{eq:partition_function_m3}
\end{eqnarray}
$F\left(\theta\right)$ is defined in (\ref{eq:define_F}). Note that
because of a relation, if $\tilde{n}_i = \tilde{n}_j$, then
$k_i = k_j$, the Jacobian of the transformation can be
written explicitly.

\subsection{Change of integral paths}
We change the integral paths in (\ref{eq:partition_function_m3}) from
$\{(-\infty,\infty)\}$ to
$\{(-\infty-\frac{Ln_{\sigma}}{2\beta\#\sigma} i,\infty -
\frac{Ln_{\sigma}}{2\beta\#\sigma} i)\}$. In what follows, `an
integral path steps over a residue' means that when an
integral path is moved, there is a residue in the region surrounded by
the initial and final integral paths. It is important to note that
each integral path in (\ref{eq:partition_function_m3}) does not step
over any residue by this change of the integral path.

Performing the change of integral paths as introduced above, we arrive at an expression of the partition function,
\begin{eqnarray}
Z_N
&=&\sum_{\theta \in \Theta\left(N\right)}
\sum_{\lambda\in \Lambda\left(\theta\right)}
J\left(\left[\lambda,\theta\right]\right)
S\left(\left[\lambda,\theta\right]\right)
\label{eq:partition_function}
\end{eqnarray}
with
\begin{eqnarray}
\!\!\!\!\!\!\!\!\!\!\!\!
J\left(\left[\lambda,\theta\right]\right)
&\equiv&
\frac{L^{N^{\prime}-\sharp \lambda}}
{N^{\prime}!}
\left(
  \prod_{\sigma^{\prime} \in g\left(\lambda,\theta\right)}
  \sharp \sigma^{\prime}
\right)
\left(
  \prod_{\sigma\in \theta}
  \sharp \sigma^{n\left(\sigma,\lambda\right)-1}
  \left(-1\right)^{\left(\sharp \sigma -1\right)}
  \left(\sharp \sigma -1\right)!
\right),
\label{eq:define_J}
\\
\!\!\!\!\!\!\!\!\!\!\!\!
S\left(\left[\lambda,\theta\right]\right)
&\equiv&
\sum_{\left\{n_{\sigma}\right\},\sigma\in\theta}
(-1)^
{
  \left(N^{\prime}-1\right)
  \sum_{\sigma\in\theta}n_{\sigma}
}
\exp
\left(
  -\frac{L^2}{4\beta}
  \sum_{\sigma\in\theta}
  \frac{n_{\delta}^2}{\#\delta}
\right)
\nonumber\\&&\times{}
\int\prod_{\sigma\in\theta}\frac{dk_{\sigma}}{2\pi}
\left(
  \prod_{\left\{\sigma,\sigma^{\prime}\right\}\in \lambda}
  \frac{d\Delta}{dk}
  \left(
    k_{\sigma}
    -k_{\sigma^{\prime}}
    -\frac{Ln_{\sigma}}{2\beta\#\sigma}i
    +\frac{Ln_{\sigma^{\prime}}}{2\beta\#\sigma^{\prime}}i
  \right)
\right)
\nonumber\\&&\times{}
\exp
\left(
  \sum_{\sigma\in\theta}
  \left(
    \begin{array}{c}
      -\beta \#\sigma k_{\sigma}^2\\
      +in_{\sigma}\sum_{\sigma^{\prime}\in\theta,\neq \sigma }
      \#\sigma
      \tilde{\Delta}
      \left(
        k_{\sigma^{\prime}}
        -k_{\sigma}
        -\frac{Ln_{\sigma}}{2\beta\#\sigma}i
        +\frac{Ln_{\sigma^{\prime}}}{2\beta\#\sigma^{\prime}}i
      \right)
    \end{array}
  \right)
\right),
\label{eq:define_S}\\ 
N^{\prime}&\equiv& \sum_{\sigma\in\theta}\#\sigma.
\label{eq:define_N} 
\end{eqnarray}

This formula enables us to calculate the partition function in any
order of the system size $L$. In other way of writing, we can calculate the partition
function for the system with finite size as
\begin{eqnarray}
Z_N& =& \sum_{n} U_{n}\left(L,\beta\right) \exp\left(-\frac{nL^2}{4N\beta}\right),\nonumber\\
& & \lim_{L\rightarrow \infty}\frac{U_n\left(L,\beta\right)}{L^N}<A(\beta),n\geq0.
\label{eq:order_form}
\end{eqnarray}
It should be remarked that $U_n$ for arbitrary $n$ is explicitly
obtained from (\ref{eq:partition_function})-(\ref{eq:define_N}).

\section{Cluster Expansion}
The cluster expansion for the equation of state is defined by
\begin{eqnarray}
p\beta & = & \lim_{L\to \infty}L^{-1}\log
             \left(
                \sum_{N=0}^{\infty}
                \left(
                   z^N
                   Z_N
                \right)
              \right)
\label{eq:grand_partition_function}                               \\
       & = & \sum_{N=1}^{\infty} z^Nb_N,                     
\end{eqnarray}
where $z$ is the fugacity (rigorously, the absolute activity),
\begin{eqnarray}
z      & = & \exp\left(\beta \mu \right).
\end{eqnarray}

The partition function $\left\{Z_N\right\}$ and the cluster integral
$\left\{b_N\right\}$ are related as
\begin{eqnarray}
b_N
&=&
\lim_{L\rightarrow\infty}
\sum_{\theta\in\Theta\left(N\right)}
\frac{\left(\#\theta-1\right)!\left(-1\right)^{\#\theta-1}\prod_{\sigma\in\theta}\#\sigma!}
{LN!}
\prod_{\sigma\in\theta}Z_{\#\sigma}.
\label{eq:cluster_integral}
\end{eqnarray}
Expanding the right hand side of eq.(\ref{eq:grand_partition_function})
in powers of $z$ and combinatorially summing up the coefficients, we can prove the relation (\ref{eq:cluster_integral}).

We define that 
\begin{eqnarray}
  \label{eq:define_lambda_c}
  \Lambda_c\left(\theta\right)
  &=&
  \left\{
    \lambda
    \left|
      \lambda\in \Lambda\left(\theta\right)\quad
      \#g\left(\lambda,\theta\right)=1
    \right.
  \right\}.
\end{eqnarray}
$\Lambda_c\left(\theta\right)$ is a set of connection pattern which consists of one cluster (this cluster is nothing to do with the cluster integral).

Substituting the expression (\ref{eq:partition_function}) of $Z_N$ into (\ref{eq:cluster_integral}), we obtain
\begin{eqnarray}
b_N
&=&
\frac{1}{L}
\sum_{\theta\in\Theta\left(N\right)}
\sum_
{
  \lambda
  \in \Lambda_c\left(\theta\right)}
J\left(\left[\lambda,\theta\right]\right)
S_{\infty}\left(\left[\lambda,\theta\right]\right),
\label{eq:cluster_integral_21}
\end{eqnarray}
where $J\left(\lambda,\theta\right)$ is defined in
(\ref{eq:define_J}), and
\begin{eqnarray}
S_{\infty}\left(\left[\lambda,\theta\right]\right)
&\equiv&
\int\prod_{\sigma\in\theta}\frac{dk_{\sigma}}{2\pi}
\left(
  \prod_{\left\{\sigma,\sigma^{\prime}\right\}\in \lambda}
  \frac{d\Delta}{dk}
  \left(
    k_{\sigma}
    -k_{\sigma^{\prime}}
  \right)
\right)
\exp
\left(
  \sum_{\sigma\in\theta}
  \left(
    -\beta \#\sigma k_{\sigma}^2
  \right)
\right).
\label{eq:define_S_inf}
\end{eqnarray}
This is proved in \ref{sec:Calculation_of_cluster_integral}.

To compare the
results in this paper with those in \cite{wadati_b}, we define a
function $K\left(k\right)$ by
\begin{eqnarray}
 \frac{d\Delta\left(k\right)}{dk}
-2\pi\delta\left(k\right)\equiv K\left(k\right).
\label{eq:define_K}
\end{eqnarray}
Note that for the non-interacting case, $K\left(k\right)\equiv$0.

In term of $K\left(k\right)$, the cluster integral $b_N$ can be expressed as follows,
\begin{eqnarray}
b_N
&=&
\frac{1}{L}
\sum_{\theta\in\Theta\left(N\right)}
\sum_
{
  \lambda
  \in \Lambda_c\left(\theta\right)}
J^{\prime}\left(\left[\lambda,\theta\right]\right)
S^{\prime}_{\infty}\left(\left[\lambda,\theta\right]\right),
\label{eq:cluster_integral_22}
\end{eqnarray}
with
\begin{eqnarray}
J^{\prime}\left(\left[\lambda,\theta\right]\right)
&\equiv&
\frac{L}
{\left(N-1\right)!}
\left(
  \prod_{\sigma\in \theta}
  \sharp \sigma^{n\left(\sigma,\lambda\right)-1}
  \left(\sharp \sigma -1\right)!
\right),
\label{eq:define_J_prime}\\
S^{\prime}_{\infty}\left(\left[\lambda,\theta\right]\right)
&\equiv&
\int\prod_{\sigma\in\theta}\frac{dk_{\sigma}}{2\pi}
\left(
  \prod_{\left\{\sigma,\sigma^{\prime}\right\}\in \lambda}
  K
  \left(
    k_{\sigma}
    -k_{\sigma^{\prime}}
  \right)
\right)
\exp
\left(
  \sum_{\sigma\in\theta}
  \left(
    -\beta \#\sigma k_{\sigma}^2
  \right)
\right).
\label{eq:define_S_prime_inf}
\end{eqnarray}
A proof of eq.(\ref{eq:cluster_integral_22}) is given also in
\ref{sec:Calculation_of_cluster_integral}.

The cluster integrals (\ref{eq:cluster_integral_22}) agree with those
derived by the TBA method (see
\ref{sec:Calculation_of_cluster_integral_with_TBA_method} for detail
calculation).  In this way, we have proved that the thermal Bethe ansatz
(TBA) method by Yang and Yang gives the exact equation of state.

\section{Concluding remarks}
Taking a one-dimensional $\delta$-function bose gas as a typical
example of integrable systems, we have derived the $N$-particle
partition function. A method in this paper, referred to as the direct
method, is an exact analysis of the partition function only based on
the periodic boundary condition.  Using the explicit expression of the partition function, we
have calculated the $N$-particle cluster integral, and proved a perfect
agreement between the results of this direct method and the thermal
Bethe ansatz (TBA) method.

The extensions and applications of the direct method to integrable and
non-integrable systems may clarify mathematical structures of the TBA method. Those problems are left for future studies.

\section*{Acknowledgment}
One of authors (G.K) likes to express his sincere thanks to T.Sasamoto, Y.Komori and A.Nishino for valuable comments and stimulating discussions.

\appendix
\section{A proof on (\ref{eq:sumrule_h})}
\label{sec:Proof_on_summation}
We prove (\ref{eq:sumrule_h}), that is,
\begin{eqnarray}
\label{eq:sumrule}
\sum_{n_1< \cdots < n_N} f\left(n_1,\cdots,n_N\right)
&=&
\frac{1}{N!}
\sum_{\theta \in \Theta\left(N\right)}
F\left(\theta\right)
\sum\left(\theta,\left\{n_i\right\}\right)
f\left(n_1,\cdots,n_N\right),
\end{eqnarray}
where $F\left(\theta\right)$ and
$\sum\left(\theta,\left\{n_i\right\}\right)$ are defined in
(\ref{eq:define_F}) and (\ref{eq:define_Sigma}), and
$f\left(n_1,\cdots,n_i,\cdots,n_j,\cdots,n_N\right)$ satisfies
\begin{eqnarray}
f\left(n_1,\cdots,n_i,\cdots,n_j,\cdots,n_N\right)
&=&
f\left(n_1,\cdots,n_j,\cdots,n_i,\cdots,n_N\right).
\end{eqnarray}

We define a semiorder on a set $\left\{\theta| \theta\subseteq\left\{1,2,\cdots\right\}\right\}$, 
\begin{eqnarray}
  \theta \leq \theta^{\prime}
  &\stackrel{{\rm def}}{\Longleftrightarrow}&
  \sigma \in \theta,\quad
  \sigma^{\prime}\in \theta^{\prime},\quad
  \sigma \subseteq \sigma^{\prime}
.
\end{eqnarray}
A sufficient condition of eq.(\ref{eq:sumrule}) is
\begin{eqnarray}
\quad\sum_{\theta^{\prime}\leq\theta}F\left(\theta^{\prime}\right)
&=&
\delta_{\theta,\theta_N}
\label{eq:sumrule_suf_1}
,\quad
 \theta \in \Theta\left(N\right),
\end{eqnarray}
where
\begin{eqnarray}
\theta_N &\equiv& \left\{\left\{1\right\},\left\{2\right\},\cdots,\left\{N\right\}\right\},\\
\delta_{\theta,\theta_N}&\equiv&
\left\{
\begin{array}{l}
0\quad {\rm if}\quad \theta = \theta_N\\
1\quad {\rm otherwise}.
\end{array}
\right.
\end{eqnarray}

We consider a function,
\begin{eqnarray}
X\left(\left\{x_{i,j}\right\}\right)&\equiv&\prod_{j=2}^{N}\left(1-\sum_{i=1}^{j-1}x_{i,j}\right).
\end{eqnarray}
and a mapping $P$,
\begin{eqnarray}
P:
h\left(\left\{x_{i,j}\right\}\right) 
\rightarrow
\sum_i n_i \theta_i \;,\quad\theta_i \in \Theta\left(N\right)
\end{eqnarray}
which satisfies the following relations
\begin{eqnarray}
P\left(h_1+h_2\right)
&=&
P\left(h_1\right)
+
P\left(h_2\right),\nonumber\\
P\left(nh\right)
&=&
nP\left(h\right),\nonumber\\
P\left(\prod_ix_{n_i,m_i}\right)
&=&
g\left(\left\{\xi|
\xi=\left\{\left\{n_i\right\},\left\{m_i\right\}\right\}\right\},\theta_N\right),
\end{eqnarray}
where $h,h_1,h_2$ are arbitrary polynomial functions, and $g\left(\lambda,\theta\right)$ is defined in (\ref{eq:define_g}).
It is readily shown that the following relation holds,
\begin{eqnarray}
P\left(X\right)
&=&\sum_{\theta\in \Theta\left(N\right)}F\left(\theta\right)\theta.
\label{eq:sumrule_1}
\end{eqnarray}
If we substitute  
\begin{eqnarray}
x_{i,j,\theta}&\equiv&\left\{
  \begin{array}{l}
    1  \quad 
    {\rm if}\;
    \sigma\in \theta \quad {\rm and} \quad n,m \in \sigma\\
    0  \quad {\rm otherwise},
  \end{array}
\right.
\end{eqnarray}
for $x_{i,j}$, the relation (\ref{eq:sumrule_1}) becomes 
\begin{eqnarray}
X\left(x_{i,j}=x_{i,j,\theta}\right)
&=&
P\left(X\left(x_{i,j}=x_{i,j,\theta}\right)\right)\nonumber\\
&=&
\sum_{\theta^{\prime} \leq \theta}
F\left(\theta^{\prime}\right)
\quad =\quad
\delta_{\theta,\theta_N}.
\label{eq:tmp_a}
\end{eqnarray}
Eq.(\ref{eq:tmp_a}) is the sufficient condition
(\ref{eq:sumrule_suf_1}) of
eq.(\ref{eq:sumrule}). Thus, (\ref{eq:sumrule_h}) is proved.

\section{A proof on (\ref{eq:jacobian_h})}
\label{sec:Proof_on_Jacobian}
We prove (\ref{eq:jacobian_h}), that is,
\begin{eqnarray}
\left|A_{\sigma,\sigma^{\prime}}\left(\theta\right)\right|
&=&
\sum_{\lambda\in \Lambda\left(\theta\right)}
\left(
  I\left(\left[\lambda,\theta\right]\right)
  \prod_{\left\{\sigma,\sigma^{\prime}\right\}\in \lambda}
  x_{\sigma,\sigma^{\prime}}
\right),
\label{eq:jacobian}
\end{eqnarray}
where
$\theta \in \Theta\left(N\right)$,
$\sigma,\sigma^\prime\in \theta$,
$x_{\sigma,\sigma^{\prime}}\equiv x_{\sigma^{\prime},\sigma}$,
$x_{\sigma,\sigma}\equiv 0$, and
 $I\left(\left[\lambda,\theta\right]\right)$ and $A_{\sigma,\sigma^{\prime}}\left(\theta\right)$ are defined in (\ref{eq:define_I}) and (\ref{eq:define_A}).

Only the terms $A_{\sigma_1,\sigma_1}\left(\theta\right)$,
$A_{\sigma_1,\sigma_2}\left(\theta\right)$,
$A_{\sigma_2,\sigma_1}\left(\theta\right)$
 and
$A_{\sigma_2,\sigma_2}\left(\theta\right)$
contain the variable $x_{\sigma_1,\sigma_2}$,
and the minor determinant becomes
\begin{eqnarray}
\left|
\begin{array}{cc}
A_{\sigma_1,\sigma_1}\left(\theta\right)                    &
A_{\sigma_1,\sigma_2}\left(\theta\right)           \\
A_{\sigma_2,\sigma_1}\left(\theta\right)           &
A_{\sigma_2,\sigma_2}\left(\theta\right)
\end{array}
\right|
&=&
\left(\alpha_{\sigma_1}+\alpha_{\sigma_2}\right)
x_{\sigma_1,\sigma_2}+\alpha_{\sigma_1}\alpha_{\sigma_2},
\label{eq:tmp_3}
\end{eqnarray}
where
\begin{eqnarray}
\alpha_{\sigma}&=&
L+\sum_{\sigma' \in \theta, \sigma' \neq \sigma_1,\sigma_2}\# \sigma'\; x_{\sigma ,\sigma'}.
\end{eqnarray}
The right hand side of (\ref{eq:tmp_3}) contains only the terms of $x_{\sigma_1,\sigma_2}$ and
$1$ when we regard the equation as a polynomial of 
$x_{\sigma_1,\sigma_2}$.  Therefore, an exponent of a variable
$x_{\sigma_1,\sigma_2}$ in the determinant (\ref{eq:jacobian}) is
1 or 0.

We define a set,
\begin{eqnarray}
\Lambda'\left( \theta \right)
&\equiv&
\left\{
  \lambda |
  \lambda = 
  \left\{
    \xi |
    \xi = \left\{\sigma , \sigma^{\prime} \right\}
    \quad \sigma , \sigma^{\prime} \in \theta 
  \right\}
\right\},\quad\theta \in \left\{\sigma\right\}.
\end{eqnarray}
While $\Lambda\left( \theta \right)$ (\ref{eq:define_Theta_tilde}) does not include $\lambda$ which
contains rings of connections, $\Lambda'\left( \theta \right)$
include such kind of $\lambda$.  In terms of this set, the determinant is written as
\begin{eqnarray}
\left|A_{\sigma,\sigma^{\prime}}\left(\theta\right)\right|
&=&
\sum_{\lambda\in \Lambda'\left(\theta\right)}
\left(
  I^{\prime}\left(\left[\lambda,\theta\right]\right)
  \prod_{\left\{\sigma,\sigma^{\prime}\right\}\in \lambda}
  x_{\sigma,\sigma^{\prime}}
\right).
\label{eq:determinant_1}
\end{eqnarray}
Here $I^{\prime}\left(\left[\lambda,\theta\right]\right)$ is an undetermined function which is projected on a polynomial of $L$. 

We make an assumption: $I^{\prime}\left(\left[\lambda,\theta\right]\right)$ is not
zero on the condition of $\lambda$ that there exists $\lambda'$ such
that
\begin{eqnarray}
  \lambda_r \subseteq \lambda\quad
  {\rm and}\quad
  \sharp
    \left\{
      \sigma
      \left|
        \xi \in \lambda^{\prime},\quad
        \sigma\in \xi
      \right.
    \right\}
    \leq
    \sharp\lambda^{\prime}.
\label{eq:condition_a}
\end{eqnarray}
This assures that the connection pattern $\lambda$ contains the rings of connections. We define a set
\begin{eqnarray}
\bar{\lambda}_r
&\equiv&
\lambda
-
\lambda_r.
\end{eqnarray}
We can choose $\lambda_r$ such that
\begin{eqnarray}
\xi\in\lambda,\quad
\xi_r\in\lambda_r\quad
\Rightarrow\quad
\xi\cap\xi_r=\emptyset.
\end{eqnarray}
This means that $\lambda_r$ consists of some cluster of $\lambda$, and  $\lambda_r$ contains some rings of connections.
In this case, it is clear that
\begin{eqnarray}
I^{\prime}\left(\left[\lambda,\theta\right]\right)
&=&
I^{\prime}\left(\left[\lambda_r,\theta\right]\right)
I^{\prime}\left(\left[\bar{\lambda}_r,\theta\right]\right)
L^{-N},
\end{eqnarray}
and that
\begin{eqnarray}
\left|
  A^{\prime}_{\sigma,\sigma^{\prime}}
  \left(\theta,\theta^{\prime}\right)
\right|
&=&
I^{\prime}\left(\left[\lambda_r,\theta\right]\right)
\prod_{\left\{\sigma,\sigma^{\prime}\right\}\in \lambda}
x_{\sigma,\sigma^{\prime}}
+
\sum_{\lambda\in \Lambda'\left(\theta\right)
,\lambda\neq\lambda_r}
\left(
  I^{\prime\prime}\left(\left[\lambda,\theta\right]\right)
  \prod_{\left\{\sigma,\sigma^{\prime}\right\}\in \lambda}
  x_{\sigma,\sigma^{\prime}}
\right),
\end{eqnarray}
where
\begin{eqnarray}
A^{\prime}_{\sigma,\sigma^{\prime}}
\left(\theta,\theta^{\prime}\right)
&=& 
\left\{
  \begin{array}{ll}
    L&
    {\rm if}\quad \sigma,\sigma^{\prime}\not\in \theta^{\prime}
    \;\sigma=\sigma^{\prime}\\
    \sum_{\sigma^{\prime\prime}\in \theta^{\prime}}
    \sharp \sigma^{\prime\prime}
    \; x_{\sigma ,\sigma^{\prime\prime}}&
    {\rm if}\quad \sigma,\sigma^{\prime}\in \theta^{\prime}
    \;\sigma =\sigma^{\prime}\\
    -\sharp \sigma^{\prime}
    \;x_{\sigma,\sigma^{\prime}} &
    {\rm if}\quad\sigma,\sigma^{\prime}\in \theta^{\prime}
    \;\sigma \neq\sigma^{\prime}\\
    0&
    {\rm otherwise},
  \end{array}
\right.
\label{eq:tmp_4}\\
&&
\theta^{\prime}
=
\left\{
  \sigma|
  \sigma\in \xi,\quad
  \xi\in \lambda_r
\right\},\quad
\sigma,\sigma^\prime\in\theta.\nonumber
\end{eqnarray}
$I^{\prime\prime}\left(\left[\lambda,\theta\right]\right)$ is
an undetermined function. 
From (\ref{eq:tmp_4}), we can see that 
\begin{eqnarray}
\sum_{\sigma\in\theta}
A^{\prime}_{\sigma,\sigma^{\prime}}
\left(\theta,\theta^{\prime}\right)
&=&0\quad {\rm if}\quad \sigma'\in \theta',
\end{eqnarray}
which indicates that row vectors of the matrix are linearly dependent; $\left|
  A^{\prime}_{\sigma,\sigma^{\prime}}
  \left(\theta,\theta^{\prime}\right) \right|$ is identically zero. This
negates the assumption that
$I^{\prime}\left(\left[\lambda,\theta\right]\right)$ is not zero on the condition
(\ref{eq:condition_a}).  Therefore, we have
\begin{eqnarray}
\left|A_{\sigma,\sigma^{\prime}}\left(\theta\right)\right|
&=&
\sum_{\lambda\in \Lambda\left(\theta\right)}
\left(
  I^{\prime}\left(\left[\lambda,\theta\right]\right)
  \prod_{\left\{\sigma,\sigma^{\prime}\right\}\in \lambda}
  x_{\sigma,\sigma^{\prime}}
\right).
\label{eq:determinant_2}
\end{eqnarray}
A difference between (\ref{eq:determinant_1}) and
(\ref{eq:determinant_2}) is the region of summation over $\lambda$.
We notice that each term of the right hand side is not made from
off-diagonal elements of
$A_{\sigma,\sigma^{\prime}}\left(\theta\right)$; all the terms
which are made from off-diagonal elements are canceled by diagonal
elements. Then, we have
\begin{eqnarray}
\left|A_{\sigma,\sigma^{\prime}}^{\prime\prime}\left(\theta\right)\right|
&=&
\sum_{\lambda\in \Lambda\left(\theta\right)}
\left(
  I^{\prime}\left(\left[\lambda,\theta\right]\right)
  \prod_{\left\{\sigma,\sigma^{\prime}\right\}\in \lambda}
  x_{\sigma,\sigma^{\prime}}
\right)
\nonumber\\
&&{}+
\sum_{\lambda\in \Lambda'\left(\theta\right)}
\sum_{\sigma_1,\sigma_2\in \theta}
I^{\left(3\right)}
\left(\left[\lambda,\theta\right],\sigma_1,\sigma_2\right)
x_{\sigma_1,\sigma_2}^2
\prod_{\left\{\sigma_3,\sigma_4\right\}\in \lambda}
x_{\sigma_3,\sigma_4}
\nonumber\\
&&{}+
\sum_{\lambda\in \Lambda'\left(\theta\right)
  ,\not\in\Lambda\left(\theta\right)}
\left(
  I^{\left(4\right)}
  \left(\left[\lambda,\theta\right]\right)
  \prod_{\left\{\sigma,\sigma^{\prime}\right\}\in \lambda}
  x_{\sigma,\sigma^{\prime}}
\right),
\label{eq:determinant_3}
\end{eqnarray}
where
\begin{eqnarray}
A_{\sigma,\sigma^{\prime}}^{\prime\prime}\left(\theta\right)
&=&
\left\{
  \begin{array}{ll}
    L+\sum_{\sigma^{\prime\prime} \in \theta}
    \sharp \sigma^{\prime\prime}
    \; x_{\sigma ,\sigma^{\prime\prime}}&
    {\rm if}\quad \sigma =\sigma^{\prime}\\
    0 &{\rm if}\quad \sigma \neq\sigma^{\prime},
  \end{array}
\right.\nonumber
\end{eqnarray}
and
$I^{\left(3\right)}\left(\left[\lambda^{\prime},\theta\right],\sigma_1,\sigma_2\right)$ and
$I^{\left(4\right)}\left(\left[\lambda^{\prime},\theta\right]\right)$
are
 undetermined functions. Therefore, we obtain
\begin{eqnarray}
I^\prime\left(\left[\lambda,\theta\right]\right)
&\equiv&
L^{N-\sharp \lambda}
\left(
  \prod_{\sigma\in \theta}
  \sharp \sigma^{n\left(\sigma,\lambda\right)-1}
\right)
\left(
  \prod_{\sigma^{\prime} \in g\left(\lambda,\theta\right)}
  \sharp \sigma^{\prime}
\right)
\nonumber\\
&=&I\left(\left[\lambda,\theta\right]\right),
\label{eq:tmp_proof_b_2}
\end{eqnarray}
and (\ref{eq:determinant_2}) with (\ref{eq:tmp_proof_b_2}) proves (\ref{eq:jacobian}).

\section{A derivation of the cluster integrals  (\ref{eq:cluster_integral_21}) and (\ref{eq:cluster_integral_22})}
\label{sec:Calculation_of_cluster_integral}
First, we prove (\ref{eq:cluster_integral_21}).
Substituting the expression of $Z_N$ in (\ref{eq:partition_function}) into (\ref{eq:cluster_integral}) yields

\begin{eqnarray}
b_N
&=&
\lim_{L\rightarrow\infty}
\sum_{\theta\in\Theta\left(N\right)}
\frac{\left(\#\theta-1\right)!\left(-1\right)^{\#\theta-1}\prod_{\sigma\in\theta}\#\sigma!}
{LN!}
\nonumber\\&&{}
\prod_{\sigma\in\theta}
\sum_{\theta^{\prime}\in\Theta\left(\#\sigma\right)}
\sum_{\lambda^{\prime}\in \Lambda\left(\theta^{\prime}\right)}
J\left(\left[\lambda^{\prime},\theta^{\prime}\right]\right)
S\left(\left[\lambda^{\prime},\theta^{\prime}\right]\right).
\label{eq:cluster_integral_1}
\end{eqnarray}
$J\left(\left[\lambda,\theta\right]\right)$ and
$S\left(\left[\lambda,\theta\right]\right)$ are defined in (\ref{eq:define_J}) and 
(\ref{eq:define_S}).
Eliminating exponentially decreasing terms with respect to $L$, we have
\begin{eqnarray}
b_N
&=&
\lim_{L\rightarrow\infty}
\sum_{\theta\in\Theta\left(N\right)}
\frac{\left(\#\theta-1\right)!\left(-1\right)^{\#\theta-1}\prod_{\sigma\in\theta}\#\sigma!}
{LN!}
\nonumber\\&&{}
\prod_{\sigma\in\theta}
\sum_{\theta^{\prime}\in\Theta\left(\sigma\right)}
\sum_{\lambda^{\prime}\in \Lambda\left(\theta^{\prime}\right)}
J\left(\left[\lambda^{\prime},\theta^{\prime}\right]\right)
S_{\infty}\left(\left[\lambda^{\prime},\theta^{\prime}\right]\right).
\label{eq:cluster_integral_2}
\end{eqnarray}
The integral $S_{\infty}\left(\lambda,\theta\right)$ is defined
in (\ref{eq:define_S_inf}).

We define
\begin{eqnarray}
\tilde{G}\left(\left[\lambda,\theta\right]\right)
&\equiv&
\left\{
    \left[\lambda',\theta'\right]
    \left|
      \begin{array}{c}
        \lambda' = \emptyset                ,\quad
        \# \theta'=1                        ,\quad
        \theta' \subset\theta               ,\quad
        \xi\in\lambda                       ,\quad
        \sigma\not\in\xi\\
        {\rm or}\\
        \lambda'\in G\left(\lambda\right)   ,\quad
        \lambda' \neq \emptyset             ,\quad
        \sigma\in\theta'                    ,\quad
        \xi\in\lambda'                      ,\quad
        \sigma\in\xi
      \end{array}
    \right.
\right\}
\end{eqnarray}
In words, $\tilde{G}\left(\left[\lambda,\theta\right]\right)$ is a set of
elements each of which consists of connection pattern $\lambda'$ and a
set $\theta'$. Here, $\lambda'$ is a cluster of the connection pattern
$\lambda$, and $\theta'$ is a subset of $\theta$ and is a set of
element which is linked with connections in $\lambda'$.
Note that 
$\tilde{G}\left(\left[\lambda,\theta\right]\right)$ contains
  ${\left[\emptyset,\left\{\sigma\right\}\right]}$ when $\sigma$ is
  not connected by $\lambda$.

From the definition, it is shown that the integral $S_{\infty}\left(\left[\lambda,\theta\right]\right)$ is factorized
into ``connected'' integrals as
\begin{eqnarray}
S_{\infty}\left(\left[\lambda,\theta\right]\right)
&=&
\prod_
{
  \left[\lambda^{\prime},\theta^{\prime}\right]
  \in \tilde{G}\left(\left[\lambda,\theta\right]\right)
}
S_{\infty}\left(\left[\lambda^{\prime},\theta^{\prime}\right]\right).
\label{eq:facorize_equation_S}
\end{eqnarray}
Similarly, the coefficient $J\left(\left[\lambda,\theta\right]\right)$ is  factorized:
\begin{eqnarray}
\left(
  \sum_{\sigma\in\theta}\#\sigma
\right)!
J\left(\left[\lambda,\theta\right]\right)
&=&
\prod_
{
  \left[\lambda^{\prime},\theta^{\prime}\right]
  \in \tilde{G}\left(\left[\lambda,\theta\right]\right)
}
\left(
  \sum_{\sigma'\in\theta'}\#\sigma'
\right)!
J_{\infty}\left(\left[\lambda^{\prime},\theta^{\prime}\right]\right).
\label{eq:facorize_equation_J}
\end{eqnarray}
Due to (\ref{eq:facorize_equation_S}) and
(\ref{eq:facorize_equation_J}), the cluster integral
(\ref{eq:cluster_integral_2}) can be rewritten as
\begin{eqnarray}
b_N
&=&
\lim_{L\rightarrow\infty}
\sum_{\theta\in\Theta\left(N\right)}
\frac{\left(\#\theta-1\right)!\left(-1\right)^{\#\theta-1}}
{LN!}
\nonumber\\&&{}
\prod_{\sigma\in\theta}
\sum_{\theta^{\prime}\in\Theta\left(\sigma\right)}
\sum_{\lambda^{\prime}\in \Lambda\left(\theta^{\prime}\right)}
\prod_
{
  \left[\lambda'',\theta''\right]
  \in \tilde{G}\left(\left[\lambda',\theta'\right]\right)
}
\left(\sum_{\sigma''\in\theta''}\#\sigma''\right)!
J\left(\left[\lambda'',\theta''\right]\right)
S_{\infty}\left(\left[\lambda'',\theta''\right]\right).
\label{eq:cluster_integral_2a}
\end{eqnarray}

\begin{figure}[p]
\begin{center}
\begin{psfrags}
      \psfrag{aa}{$\theta$}
      \psfrag{ab}{$\left[\lambda,\theta\right]$}
      \psfrag{aca}{$\left[\lambda'_1,\theta'_1\right]$}
      \psfrag{acb}{$\left[\lambda'_2,\theta'_2\right]$}
      \psfrag{acc}{$\left[\lambda'_3,\theta'_3\right]$}
      \psfrag{ba}{$\Lambda\left(\theta\right)$}
  \psfig{file=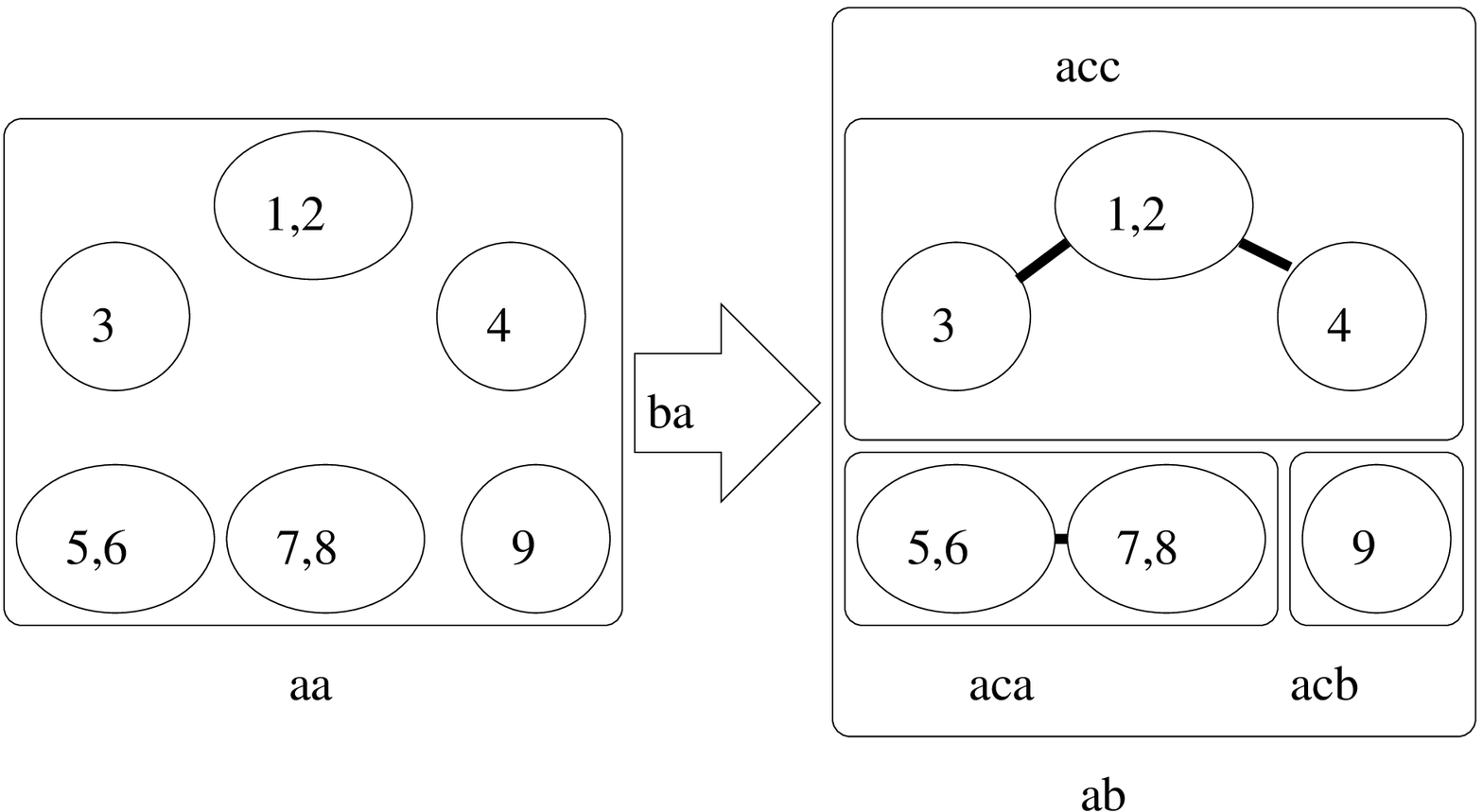 , scale = 0.50}
\end{psfrags}
\caption{A graphical representation of one of patterns which are summed up in the light hand side of eq.(\ref{eq:combinatorial_c1}).}
\label{fig:combi_1l}
\begin{psfrags}
      \psfrag{aa}{$\theta''$}
      \psfrag{aba}{$\theta'_1$}
      \psfrag{abb}{$\theta'_2$}
      \psfrag{abc}{$\theta'_3$}
      \psfrag{aca}{$\left[\lambda'_1,\theta'_1\right]$}
      \psfrag{acb}{$\left[\lambda'_2,\theta'_2\right]$}
      \psfrag{acc}{$\left[\lambda'_3,\theta'_3\right]$}
      \psfrag{ba}{$\Theta\left(\sigma''\right)$}
      \psfrag{bb}{$\Lambda_c\left(\theta'\right)$}
  \psfig{file=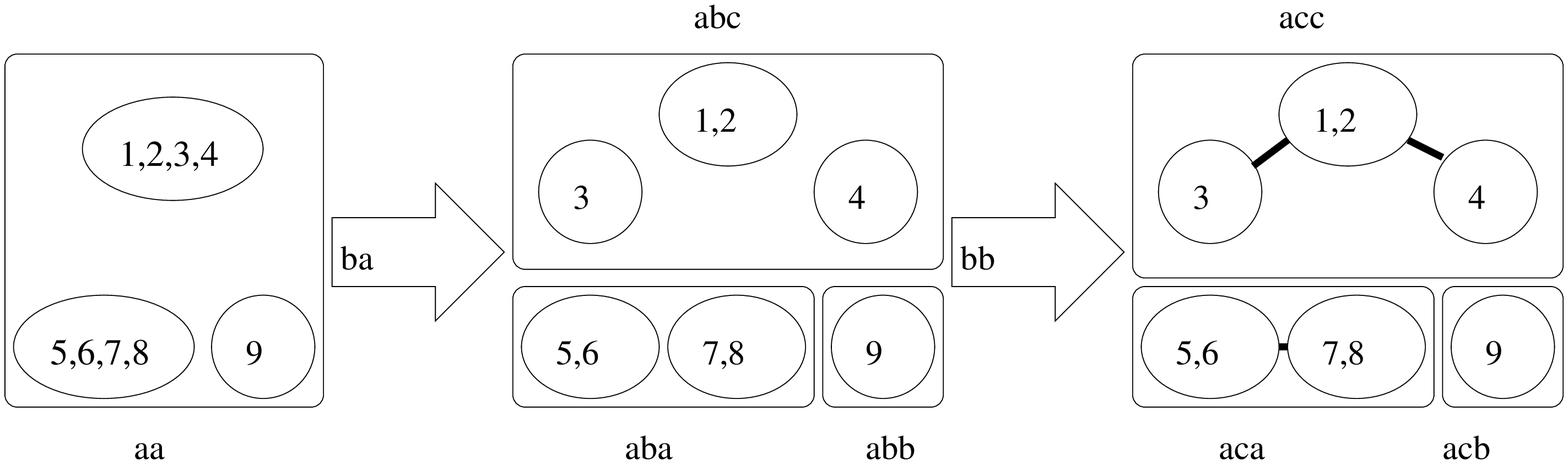 , scale = 0.50}
\end{psfrags}
\caption{A graphical representation of one of patterns which are summed up in the right hand side of eq.(\ref{eq:combinatorial_c1}).}
\label{fig:combi_1r}
\end{center}
\end{figure}

We can show that for arbitrary function $f\left(\left[\lambda,\theta\right]\right)$,
\begin{eqnarray}
\sum_{\theta\in\Theta\left(N\right)}
\sum_{\lambda\in \Lambda\left(\theta\right)}
\prod_
{
  \left[\lambda^{\prime},\theta^{\prime}\right]
  \in \tilde{G}\left(\left[\lambda,\theta\right]\right)
}
f\left(\left[\lambda^{\prime},\theta^{\prime}\right]\right)
&=&
\sum_{\theta''\in\Theta\left(N\right)}
\prod_{\sigma''\in\theta''}
\sum_{\theta'\in\Theta\left(\sigma''\right)}
\sum_
{
  \lambda^{\prime}
  \in \Lambda_c\left(\theta'\right)}
  f\left(\left[\lambda^{\prime},\theta^{\prime}\right]\right).
\label{eq:combinatorial_c1}
\end{eqnarray}
The left hand side of (\ref{eq:combinatorial_c1}) is a summation over
 all the patterns which are generated by the following process:
first divide a set $\left\{1,\cdots,N\right\}$ into elements of
$\theta$,
then connect them with $\lambda$.
The right hand side of (\ref{eq:combinatorial_c1}) is a summation over
all the patterns which generated by the following process:
first divides a set $\left\{1,\cdots,N\right\}$ into sets of $\theta''$
each of which is a direct sum of elements connected by a cluster
$\lambda'$,
second divide each of the sets into elements of $\theta'$, then define
connection pattern $\lambda'$, which consists of one cluster, of elements
in $\theta'$.
Fig.\ref{fig:combi_1l} and Fig.\ref{fig:combi_1r} illustrate graphical representations of both sides of patterns in (\ref{eq:combinatorial_c1}).

Due to (\ref{eq:combinatorial_c1}), the cluster integral
(\ref{eq:cluster_integral_2a}) becomes
\begin{eqnarray}
b_N
&=&
\lim_{L\rightarrow\infty}
\sum_{\theta\in\Theta\left(N\right)}
\frac{\left(\#\theta-1\right)!\left(-1\right)^{\#\theta-1}
}
{LN!}
\prod_{\sigma\in\theta}
\sum_{\theta^{\prime}\in\Theta\left(\sigma\right)}
\prod_{\sigma^{\prime}\in\theta^{\prime}}
\#\sigma^{\prime}!
\nonumber\\&&{}
\sum_{\theta^{\prime\prime}\in\Theta\left(\sigma^{\prime}\right)}
\sum_
{
  \lambda^{\prime\prime}
  \in \Lambda_c\left(\theta^{\prime\prime}\right)
}
J\left(\left[\lambda'',\theta''\right]\right)
S_{\infty}\left(\left[\lambda'',\theta''\right]\right).
\label{eq:cluster_integral_3}
\end{eqnarray}

We define a family of sets $\tilde{g}\left(\tau\right)$,
\begin{eqnarray}
  \tilde{g}\left(\tau\right)
  &=&
  \left\{
    \sigma
    \left|
      \sigma=\bigcup_{\sigma'\in\theta'}\sigma'\quad
      \theta'\in\tau
    \right.
  \right\},
\end{eqnarray}
where an element of $\tau$
is also a family of sets. To repeat, an element of $\tilde{g}\left(\tau\right)$
is a sum of sets of which a family of sets, an element of $\tau$,
consists. For example,
\begin{eqnarray}
  \tilde{g}
  \left(
    \left\{
      \left\{
        \left\{1,2\right\},
        \left\{3\right\}
      \right\},
      \left\{
        \left\{4,5\right\},
        \left\{6,7\right\}
      \right\}
    \right\}
  \right)
  &=&
  \left\{
    \left\{1,2,3\right\}
    \left\{4,5,6,7\right\}
  \right\}.
\end{eqnarray}

\begin{figure}[p]
\begin{center}
\begin{psfrags}
      \psfrag{aa}{$\theta''$}
      \psfrag{ab}{$\tau$}
      \psfrag{ba}{$\Theta\left(\theta''\right)$}
  \psfig{file=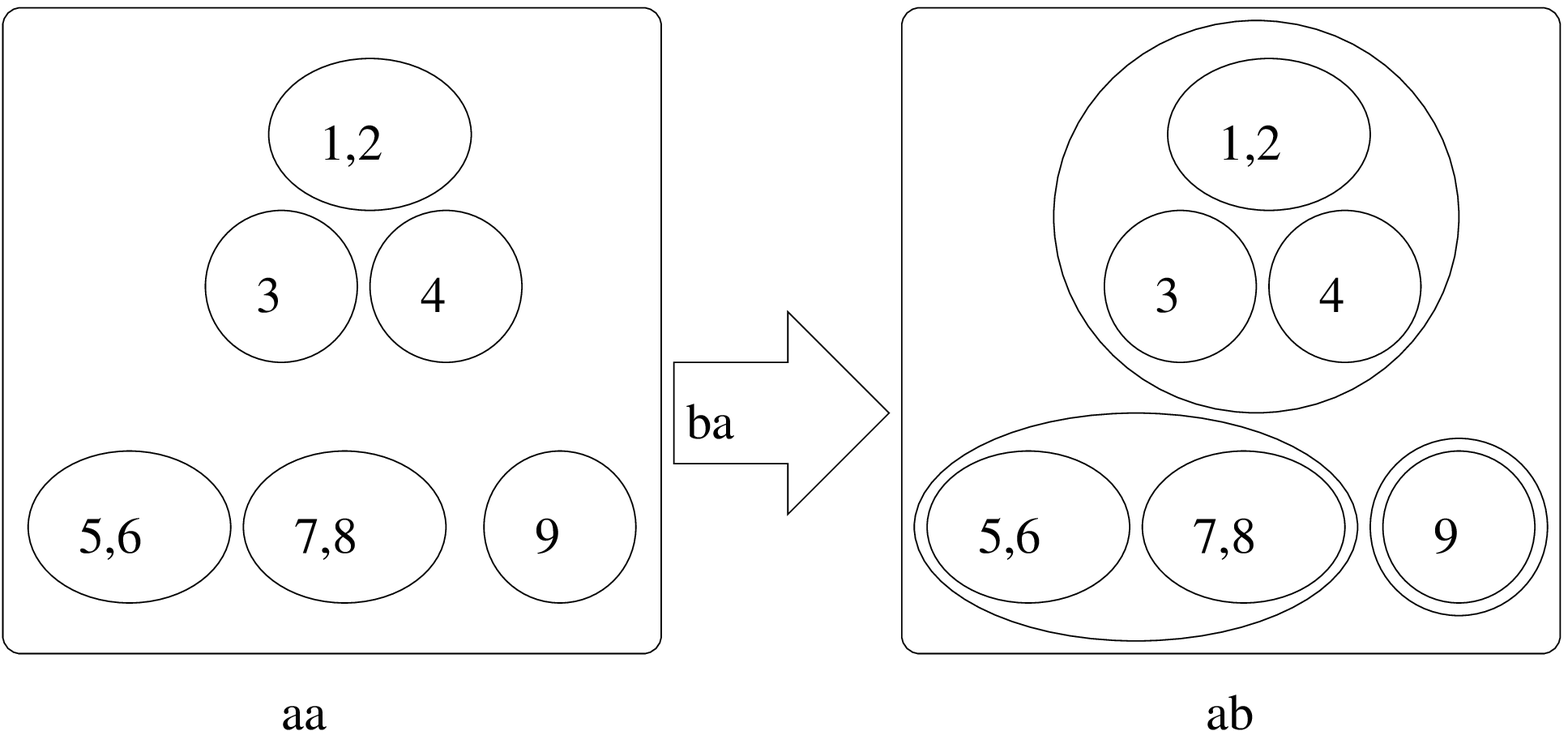 , scale = 0.50}
\end{psfrags}
\caption{A graphical representation of one of patterns which are summed up in the left hand side of eq.(\ref{eq:combinatorial_c2}).}
\label{fig:combi_2l}
\begin{psfrags}
      \psfrag{aa}{$\theta$}
      \psfrag{aba}{$\theta'_1$}
      \psfrag{abb}{$\theta'_2$}
      \psfrag{abc}{$\theta'_3$}
      \psfrag{ba}{$\Theta\left(\sigma\right)$}
  \psfig{file=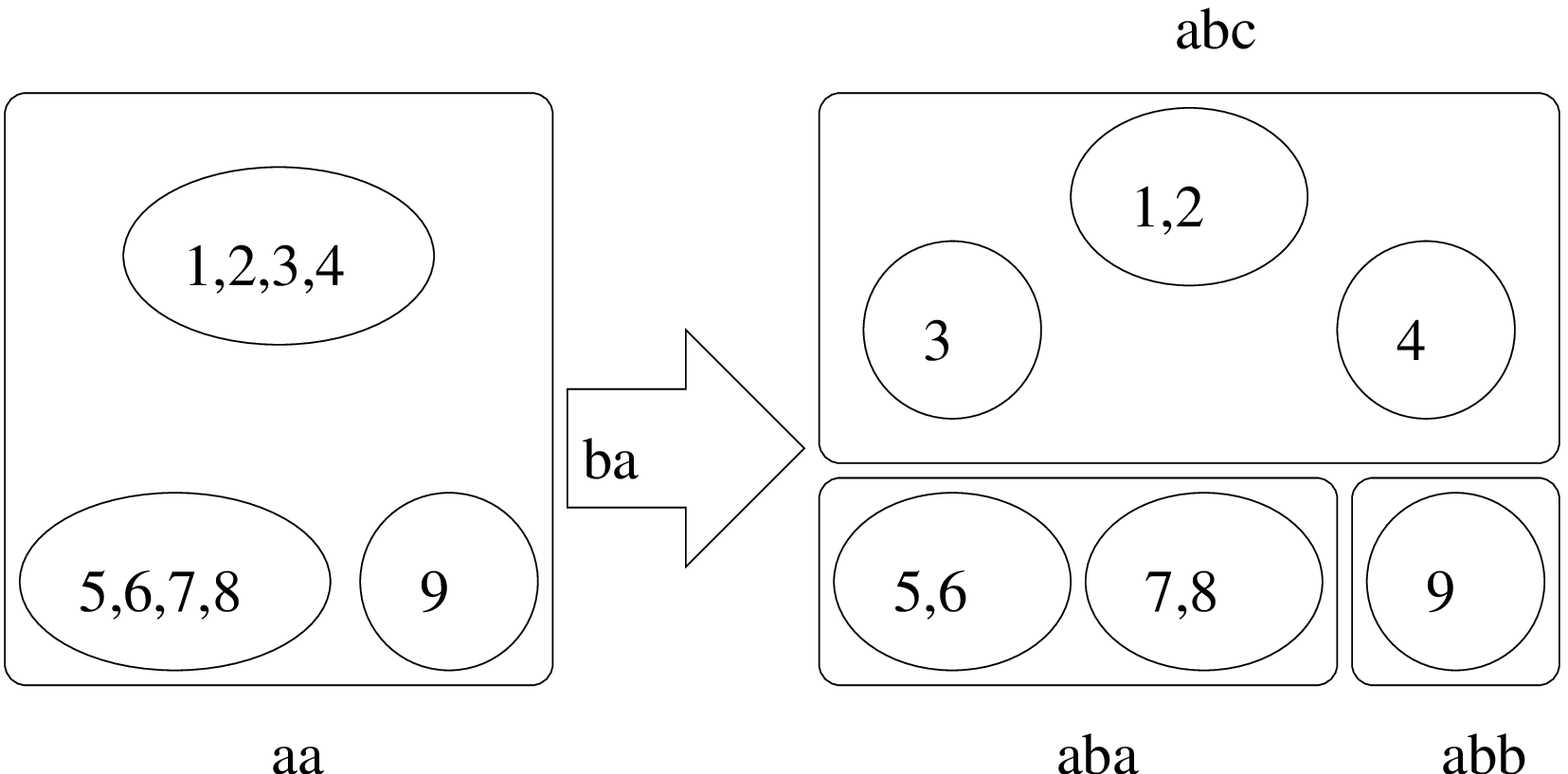 , scale = 0.50}
\end{psfrags}
\caption{A graphical representation of one of patterns which are summed up in the right hand side of eq.(\ref{eq:combinatorial_c2}).}
\label{fig:combi_2r}
\end{center}
\end{figure}

We can show that for arbitrary functions $f_1\left(\theta\right)$,
$f_2\left(\sigma\right)$,
\begin{eqnarray}
\sum_{\theta''\in\Theta\left(N\right)}
\left(
  \prod_{\sigma'\in\theta''}
  f_2\left(\sigma'\right)
\right)
\sum_{\tau\in\Theta\left(\theta''\right)}
f_1\left(\tilde{g}\left(\tau\right)\right)
&=&
\sum_{\theta\in\Theta\left(N\right)}
f_1\left(\theta\right)
\prod_{\sigma\in\theta}
\sum_{\theta'\in\Theta\left(\sigma\right)}
\prod_{\sigma'\in\theta'}
f_2\left(\sigma'\right).
\label{eq:combinatorial_c2}
\end{eqnarray}
In the left hand side of (\ref{eq:combinatorial_c2}), one sums up all
the patterns which are generated by the following process:
first divide $\left\{1,\cdots,N\right\}$ into elements in $\theta''$,
then make family of sets $\tau$.
Each of the sets consist of some elements in $\theta''$, and
$\tilde{g}\left(\tau\right)=\theta$.
In the right hand side of (\ref{eq:combinatorial_c2}), one sums up all
the patterns which are generated by the following process:
first divide a set $\left\{1,\cdots,N\right\}$ into elements of
$\theta$,
then divide each of elements in $\theta$ into elements of
$\theta'$
Fig.\ref{fig:combi_2l} and Fig.\ref{fig:combi_2r} illustrate graphical
representations of both sides of patterns in
(\ref{eq:combinatorial_c2}).

Using a formula (\ref{eq:combinatorial_c2}) in
(\ref{eq:cluster_integral_2a}), $b_N$ becomes
\begin{eqnarray}
b_N
&=&
\lim_{L\rightarrow\infty}
\sum_{\theta\in\Theta\left(N\right)}
\left(
  \prod_{\sigma\in\theta}
  \#\sigma!
  \sum_{\theta^{\prime\prime}\in\Theta\left(\sigma\right)}
  \sum_
  {
    \lambda^{\prime\prime}
    \in \Lambda_c\left(\theta^{\prime\prime}\right)}
  J\left(\left[\lambda'',\theta''\right]\right)
  S_{\infty}\left(\left[\lambda'',\theta''\right]\right).
\right)
\nonumber\\&&{}
\sum_{\tau\in\Theta\left(\theta\right)}
\frac{\left(\#\tau-1\right)!\left(-1\right)^{\#\tau-1}
}
{LN!}
\label{eq:cluster_integral_4}
\end{eqnarray}
Note that we have used a relation $\#\tilde{g}\left(\tau\right)=\#\tau$.

Substitutions of $\frac{L^N}{N!}$ into $Z_N$
in (\ref{eq:cluster_integral}) and 
(\ref{eq:grand_partition_function}) give
a relation,
\begin{eqnarray}
\sum_{\theta\in\Theta\left(N\right)}
\left(\#\theta-1\right)!\left(-1\right)^{\#\theta-1}
&=&
\delta_{N,1}.
\label{eq:sum_relation_1}
\end{eqnarray}
Using this relation in (\ref{eq:cluster_integral_4}), we finally obtain
\begin{eqnarray}
b_N
&=&
\frac{1}{L}
\sum_{\theta\in\Theta\left(N\right)}
\sum_
{
  \lambda
  \in \Lambda_c\left(\theta\right)}
J\left(\left[\lambda,\theta\right]\right)
S_{\infty}\left(\left[\lambda,\theta\right]\right),
\label{eq:cluster_integral_5}
\end{eqnarray}
which is (\ref{eq:cluster_integral_21}).
Note that the right hand side of eq.(\ref{eq:cluster_integral_5}) does
not depend on $L$. Therefore, we do not write $\lim_{L\rightarrow\infty}$ any more.

In the no-interaction limit, where
$\frac{d\Delta}{dk}\left(k\right)$ is replaced with $\delta\left(k\right)$, eq.(\ref{eq:cluster_integral_5}) gives
\begin{eqnarray}
\frac{1}{N}
&=&
\sum_{\theta\in\Theta\left(N\right)}
\sum_
{\lambda\in \Lambda_c\left(\theta\right)}
\frac{1}
{\left(N-1\right)!}
\left(
  \prod_{\sigma\in \theta}
  \sharp \sigma^{n\left(\sigma,\lambda\right)-1}
  \left(-1\right)^{\left(\sharp \sigma -1\right)}
  \left(\sharp \sigma -1\right)!
\right).
\label{eq:sum_relation_2}
\end{eqnarray}

\begin{figure}[p]
\begin{center}
\begin{psfrags}
      \psfrag{a}{$\lambda_9$}
  \psfig{file=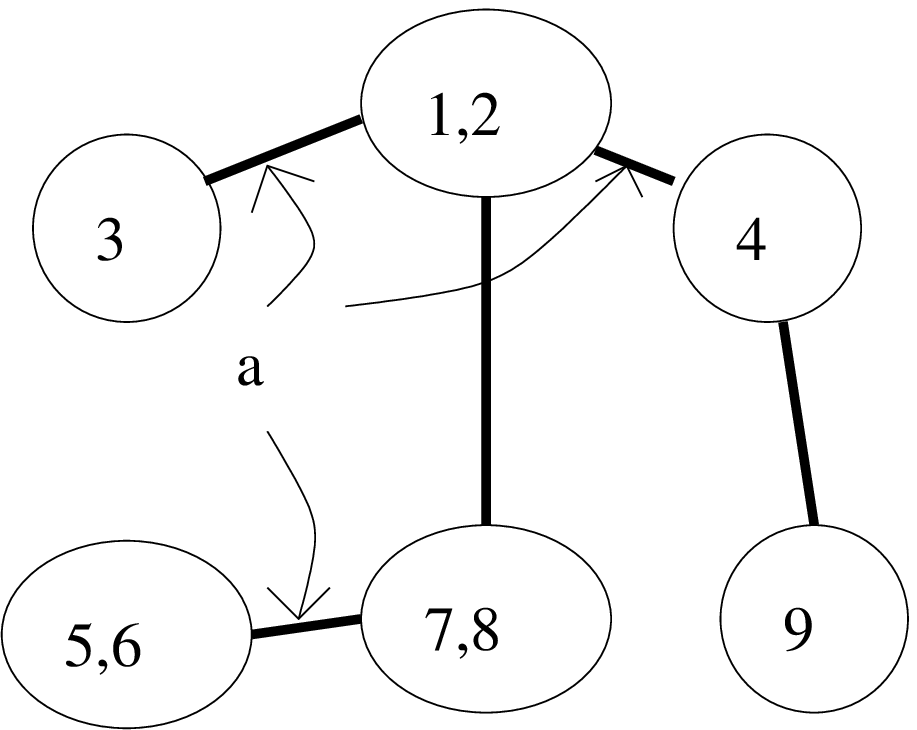 , scale = 0.55}
\end{psfrags}
\caption{A graphical representation of $\left[\lambda''_9,\theta_9\right]$ in (\ref{eq:tmp_proof_c_1}).}
\label{fig:proof_c1}
\begin{psfrags}
  \psfig{file=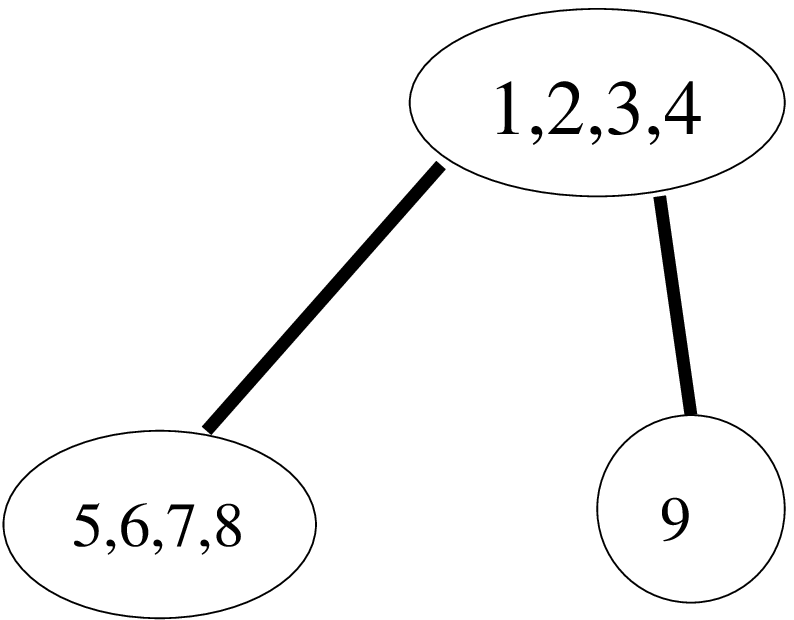 , scale = 0.55}
\end{psfrags}
\caption{A graphical representation of $\left[\lambda'_9,\theta'_9\right]$ in (\ref{eq:tmp_proof_c_1}).}
\label{fig:proof_c2}
\end{center}
\end{figure}

Next, we prove (\ref{eq:cluster_integral_22}).
We define
\begin{eqnarray}
  \label{eq:define_D}
  D\left(\left[\lambda,\theta\right],\lambda'\right)
  &=&
  [\lambda'',\theta'']\nonumber\\
  &\equiv&
  \left[
  \left\{
    \left\{
      \sigma_1,
      \sigma_2
    \right\}
    \left|
      \begin{array}{c}
        \sigma_1,\sigma_2\in\theta''\\
        \sigma_1\in \sigma_3\quad
        \sigma_2\in \sigma_4\\
        \left\{
          \sigma_3,
          \sigma_4
        \right\}
        \in\lambda,\not\in\lambda''
      \end{array}
    \right.
  \right\}
  ,g\left(\left[\lambda',\theta\right]\right)
  \right]\\
  &&\lambda'\subseteq\lambda\nonumber
\end{eqnarray}
$\theta''$ is a set of a direct sum of
elements linked with one cluster in $\lambda'$, and $\lambda''$
is a connection pattern of a set $\theta''$. A connection in $\lambda''$
links two elements of $\theta''$, where elements of $\theta$, each
of which is a subset of one of the two elements, are connected by
$\lambda$. For example,
\begin{eqnarray}
  D\
  \left(
    \left[\lambda_9'',\theta_9\right],\lambda_9
  \right)
  &=&
  \left[\lambda'_9,\theta'_9\right],
\label{eq:tmp_proof_c_1}
\end{eqnarray}
where
\begin{eqnarray}
  \lambda_{9}
  &=&
  \left\{
    \left\{
      \left\{1,2\right\},
      \left\{3\right\}
    \right\}
    ,
    \left\{
      \left\{1,2\right\},
      \left\{4\right\}
    \right\}
    ,
    \left\{
      \left\{5,6\right\},
      \left\{7,8\right\}
    \right\}
  \right\}
  \;,\\
  \theta_{9}
  &=&
  \left\{
    \left\{1,2\right\},
    \left\{3\right\},
    \left\{4\right\},
    \left\{5,6\right\},
    \left\{7,8\right\},
    \left\{9\right\}
  \right\} \;,\\
    \lambda''_9
    &=&
    \left\{
\begin{array}{c}
      \left\{\left\{1,2\right\},\left\{  3\right\}\right\},
      \left\{\left\{1,2\right\},\left\{  4\right\}\right\},
      \left\{\left\{1,2\right\},\left\{7,8\right\}\right\},\\
      \left\{\left\{  4\right\},\left\{  9\right\}\right\},
      \left\{\left\{5,6\right\},\left\{7,8\right\}\right\}
\end{array}
    \right\}\;,\\
    \lambda'_9
    &=&
    \left\{
      \left\{\left\{1,2,3,4\right\},\left\{5,6,7,8\right\}\right\},
      \left\{\left\{1,2,3,4\right\},\left\{9\right\}\right\}
    \right\}\;,\\
    \theta'_9
    &=&
    \left\{
      \left\{1,2,3,4\right\},
      \left\{5,6,7,8\right\},
      \left\{9\right\}
    \right\}.
\end{eqnarray}
Fig.\ref{fig:proof_c1} and Fig.\ref{fig:proof_c2} illustrate $\left[\lambda''_9,\theta_9\right]$ and $\left[\lambda'_9,\theta'_9\right]$ in (\ref{eq:tmp_proof_c_1}).

By substituting $K \left(k\right)+\delta\left(k\right)$ into
$\frac{d\Delta}{dk}\left(k\right)$, (\ref{eq:cluster_integral_5}) is
changed into
\begin{eqnarray}
  b_N
  &=&
  \frac{1}{L}
  \sum_{\theta\in\Theta\left(N\right)}
  \sum_
  {
    \lambda
    \in \Lambda_c\left(\theta\right)}
  J\left(\left[\lambda,\theta\right]\right)
  \sum_{\lambda'\subseteq\lambda}
  S^{\prime}_{\infty}
  \left(
    D\left(\left[\lambda,\theta\right],\lambda'\right)
  \right),\nonumber
\end{eqnarray}
where $S^{\prime}_{\infty}\left(\left[\lambda,\theta\right]\right)$ is defined in (\ref{eq:define_S_prime_inf}).
With the explicit expression of $J\left(\left[\lambda,\theta\right]\right)$,
we have
\begin{eqnarray}
  b_N
  &=&
  \frac{1}{\left(N-1\right)!}
  \sum_{\theta\in\Theta\left(N\right)}
  \sum_
  {
    \lambda
    \in \Lambda_c\left(\theta\right)}
  \sum_{\lambda'\subseteq\lambda}
  S^{\prime}_{\infty}
  \left(
    D\left(\left[\lambda,\theta\right],\lambda'\right)
  \right)
\nonumber\\&&
  \prod_{\sigma\in\theta}
  \#\sigma^{n\left(\sigma,\lambda\right)-1}
\left(-1\right)^{\left(\#\sigma-1\right)}
\left(\#\sigma-1\right)!.
  \label{eq:sum_relation_11_a}
\end{eqnarray}

\begin{figure}[p]
\begin{center}
\begin{psfrags}
      \psfrag{aa}{$\theta$}
      \psfrag{ab}{$\left[\lambda,\theta\right]$}
      \psfrag{ac}{$D\left(\left[\lambda,\theta\right],\lambda''\right)=\left[\lambda',\theta'\right]$}
      \psfrag{ba}{$\Lambda\left(\theta\right)$}
      \psfrag{bb}{{\small $D\left(\left[\lambda,\theta\right],\lambda''\right)$}}
      \psfrag{c}{$\lambda''$}
  \psfig{file=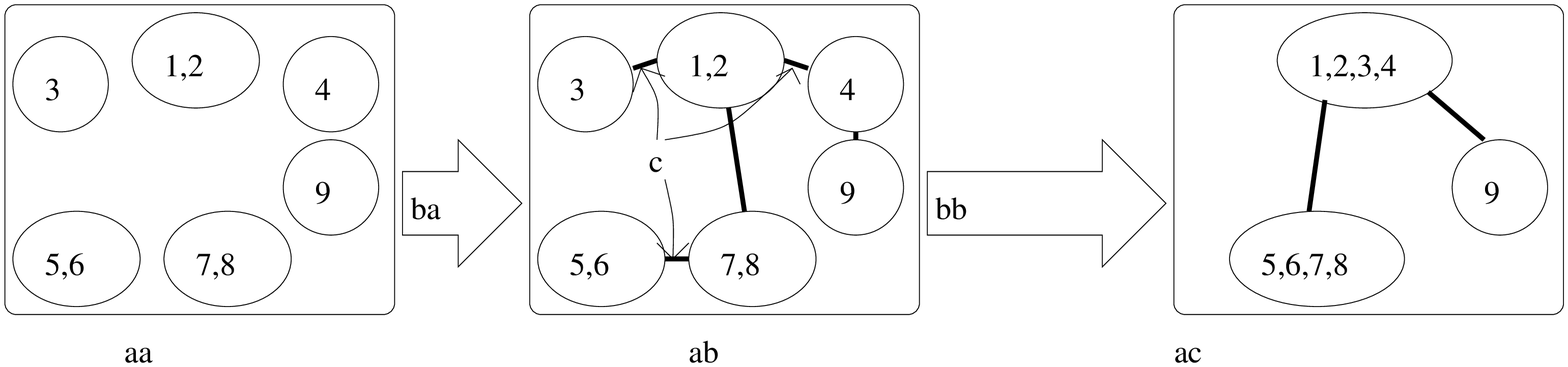 , scale = 0.50}
\end{psfrags}
\caption{A graphical representation of one of patterns which are summed up in the left hand side of eq.(\ref{eq:combinatorial_c3}).}
\label{fig:combi_3l}
\begin{psfrags}
      \psfrag{aa}{$\theta'$}
      \psfrag{ab}{$\left[\lambda',\theta'\right]$}
      \psfrag{aca}{$\theta''_1$}
      \psfrag{acb}{$\theta''_2$}
      \psfrag{acc}{$\theta''_3$}
      \psfrag{ada}{$\left[\lambda''_1,\theta''_1\right]$}
      \psfrag{adb}{$\left[\lambda''_2,\theta''_2\right]$}
      \psfrag{adc}{$\left[\lambda''_3,\theta''_3\right]$}
      \psfrag{ba}{$\Lambda\left(\theta'\right)$}
      \psfrag{bb}{$\Theta\left(\sigma'\right)$}
      \psfrag{bc}{$\Lambda\left(\theta''\right)$}
      \psfrag{bd}{$\sigma_1\cdots$}
  \psfig{file=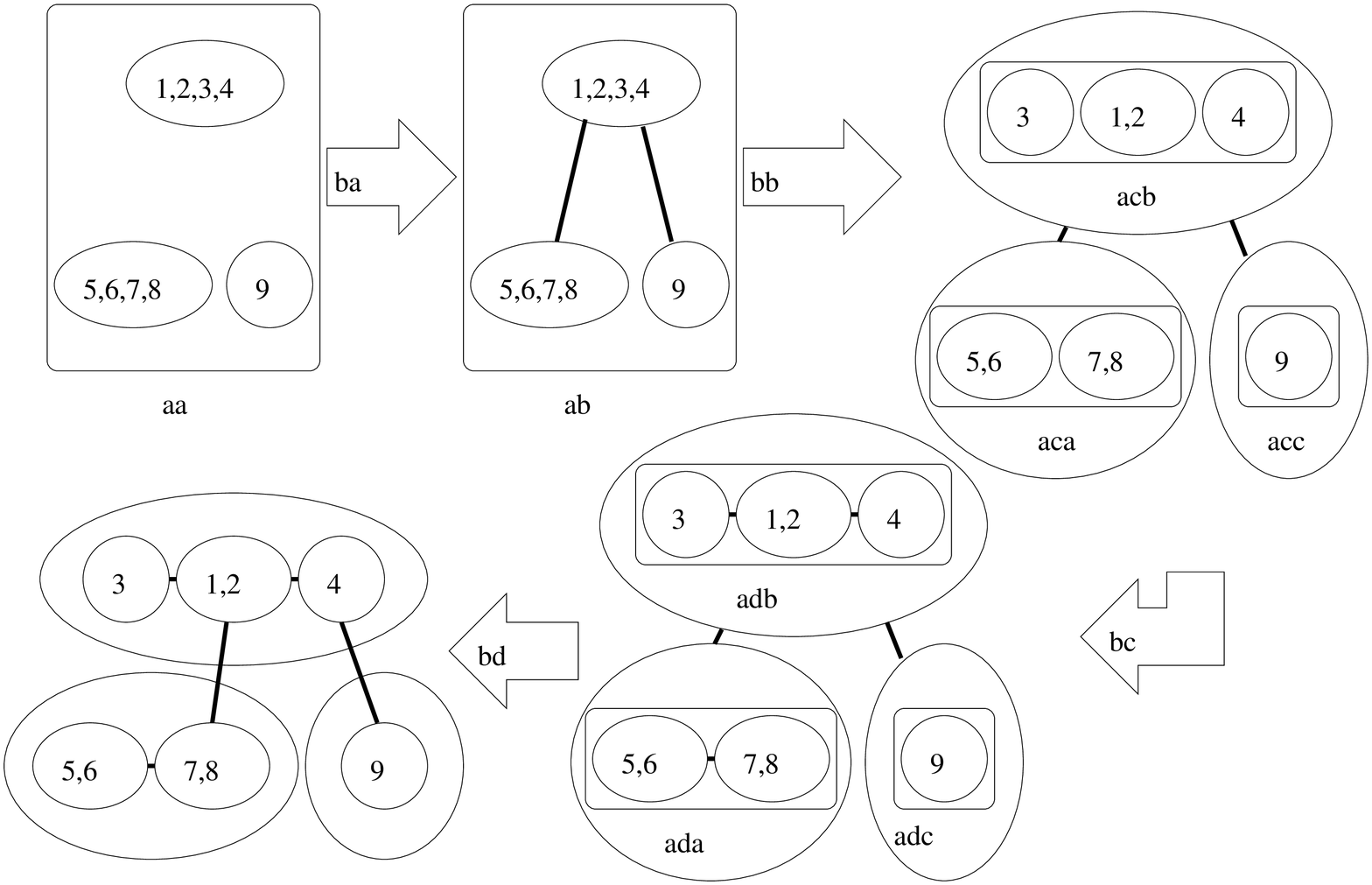 , scale = 0.50}
\end{psfrags}
\caption{A graphical representation of one of  patterns which are summed up in the right hand side of eq.(\ref{eq:combinatorial_c3}).}
\label{fig:combi_3r}
\end{center}
\end{figure}

We can show that for arbitrary functions $f_1\left(n,\sigma\right)$ and
$f_2\left(\left[\lambda,\theta\right]\right)$,
\begin{eqnarray}
&&\sum_{\theta\in\Theta\left(N\right)}
  \sum_{\lambda\in \Lambda_c\left(\theta\right)}
  \sum_{\lambda''\subseteq\lambda}
  f_2
  \left(
    D\left(\left[\lambda,\theta\right],\lambda''\right)
  \right)
  \prod_{\sigma\in\theta}
  f_1\left(n\left(\sigma,\lambda\right),\sigma\right)\nonumber\\
&=&
  \sum_{\theta'\in\Theta\left(N\right)}
  \sum_{\lambda'\in \Lambda_c\left(\theta'\right)}
  f_2\left(\left[\lambda',\theta'\right]\right)
  \prod_{\sigma'\in\theta'}
  \sum_{\theta''\in\Theta\left(\sigma'\right)}
  \sum_{\lambda''\in \Lambda_c\left(\theta''\right)}
\nonumber\\&&
  \sum_{\sigma_1,\cdots,\sigma_{n\left(\sigma,\lambda\right)}\in\theta''}
  \prod_{\sigma\in\theta''}
  f_1\left(n\left(\sigma,\lambda''\right)
    +\sum_{i=1}^{n\left(\sigma,\lambda\right)}\delta_{\sigma,\sigma_i},\sigma\right).
  \label{eq:combinatorial_c3}
\end{eqnarray}
The meaning of the left hand side of (\ref{eq:combinatorial_c3}) is to
sum up all the patterns generated by the following process:
first, one divides a set into elements of $\theta$,
second defines a connection pattern $\lambda$,
then makes $\theta'$ and $\lambda'$ to join elements of $\theta$
together which are connected with $\lambda''$.
The meaning of the right hand side of (\ref{eq:combinatorial_c3}) is
to sum up all the patterns generated by the following process:
first one divides a set into elements of $\theta'$,
second makes a connection pattern $\lambda'$,
third divides each elements of the set $\theta'$ into elements of
$\theta''$,
fourth defines connection pattern $\lambda''$ on elements of $\theta''$
each of which is a subset of an element in $\theta'$,
then connects elements each of which is a subset of one of two elements
in $\theta'$ linked with a connection pattern $\lambda'$.
Fig.\ref{fig:combi_3l} and Fig.\ref{fig:combi_3r} show graphical representations of both sides of patterns in (\ref{eq:combinatorial_c3}).
(\ref{eq:combinatorial_c3}) is similar to
(\ref{eq:combinatorial_c2}), in the sense that if
$f_1\left(n,\sigma\right)$ and
$f_2\left(\left[\lambda,\theta\right]\right)$ in
eq.(\ref{eq:combinatorial_c3}) are independent of $n$ and $\lambda$;
there is no connection pattern dependence, then
eq.(\ref{eq:combinatorial_c3}) becomes (\ref{eq:combinatorial_c2}).

By use of  (\ref{eq:combinatorial_c3}), we rewrite (\ref{eq:sum_relation_11_a}) as
\begin{eqnarray}
b_N
&=&
\frac{1}{\left(N-1\right)!}
\sum_{\theta\in\Theta\left(N\right)}
\sum_
{
  \lambda
  \in \Lambda_c\left(\theta\right)}
S^{\prime}_{\infty}\left(\left[\lambda,\theta\right]\right)
\prod_{\sigma\in\theta}
\sum_{\theta'\in \Theta\left(\sigma\right)}
  \sum_{\lambda'\in \Lambda_c\left(\theta'\right)}
\nonumber\\&&{}
\sum_{\sigma'_1,\cdots,\sigma'_{n\left(\lambda,\theta\right)}\in\theta'}
\prod_{\sigma'\in\theta'}
  \sharp\sigma^{\prime n\left(\sigma',\lambda'\right)+\sum_{i=1}^{n\left(\sigma,\lambda\right)}\delta_{\sigma',\sigma'_i}-1}
  \left(-1\right)^{\#\sigma'-1}
  \left(\sharp \sigma' -1\right)!.
\nonumber
\end{eqnarray}
Since;
\begin{eqnarray}
\sum_{\sigma_1,\cdots,\sigma_{n}\in\theta}
\prod_{\sigma\in\theta}
  \sharp\sigma^{ \sum_{i=1}^{n}\delta_{\sigma,\sigma_i}}
&=&
\left(
  \sum_{\sigma\in \theta}  \sharp \sigma
\right)^{n}
,
\end{eqnarray}
$b_N$ becomes
\begin{eqnarray}
b_N
&=&
\sum_{\theta\in\Theta\left(N\right)}
\sum_
{
  \lambda
  \in \Lambda_c\left(\theta\right)}
S^{\prime}_{\infty}\left(\left[\lambda,\theta\right]\right)
\frac{1}{\left(N-1\right)!}
\prod_{\sigma\in\theta}
  \sharp \sigma^{n\left(\sigma,\lambda\right)}
\nonumber\\&&{}
\sum_{\theta^{\prime}\in \Theta\left(\sigma\right)}
\sum_{
  \lambda^{\prime}
  \in \Lambda_c\left(\theta^{\prime}\right)
}
\left(
\prod_{\sigma^{\prime}\in \theta^{\prime}}
  \sharp {\sigma^{\prime}}^{n\left(\sigma^{\prime},\lambda^{\prime}\right)-1}
  \left(-1\right)^{\#\sigma^{\prime}-1}
  \left(\sharp \sigma^{\prime} -1\right)!
\right).
\label{eq:sum_relation_11_b}
\end{eqnarray}
Using the relation(\ref{eq:sum_relation_2}), we arrive at (\ref{eq:cluster_integral_22}),
\begin{eqnarray}
b_N
&=&
\frac{1}{L}
\sum_{\theta\in\Theta\left(N\right)}
\sum_
{
  \lambda
  \in \Lambda_c\left(\theta\right)}
J^{\prime}\left(\left[\lambda,\theta\right]\right)
S^{\prime}_{\infty}\left(\left[\lambda,\theta\right]\right),
\end{eqnarray}
where $J^{\prime}\left(\left[\lambda,\theta\right]\right)$ is defined in (\ref{eq:define_J_prime}).

\section{The cluster integral (\ref{eq:cluster_integral_22}) by use of the TBA}
\label{sec:Calculation_of_cluster_integral_with_TBA_method}
A result of the bosonic formulation of the TBA~\cite{wadati_b} is
\begin{eqnarray}
p\beta
&=&
-\int\frac{dk}{2\pi}
\log\left(
  1
  -z\exp\left(
    -\beta\left(
      k^2+\pi\left(k\right)
    \right)
  \right)
\right),
\label{eq:integrant_of_grand_partition_function}
\end{eqnarray}
where
\begin{eqnarray}
\pi\left(k\right)
&=&
\frac{1}{\beta}
\int\frac{dq}{2\pi}
K\left(k-q\right)
\log\left(
  1
  -z\exp\left(
    -\beta\left(
      q^2+\pi\left(q\right)
    \right)
  \right)
\right).
\label{eq:self_enargy}
\end{eqnarray}
We define a function,
\begin{eqnarray}
F_N\left(k\right)
&\equiv&
\frac{1}{N!}\left.\frac{\partial^N\exp\left(-\beta\pi\left(k\right)\right)}{\partial z^N}\right|_{z=0},
\end{eqnarray}
and prove that
\begin{eqnarray}
F_N\left(k\right)
&=&
\frac{1}{N!}
\sum_{\theta\in\Theta\left(N\right)}
\sum_
{
  \lambda\in\Lambda_c\left(\theta_{+}\right)
}
\left(
  \prod_{\sigma\in\theta,\sigma \neq \left\{0\right\}}
  \left(\#\sigma-1\right)!
  \#\sigma^{n\left(\sigma,\lambda\right)-1}
\right)
\tilde{S}\left(\left[\lambda,\theta\right],k,\left\{0\right\}\right)
\label{eq:expansion_of_self_enargy}
,\quad\quad \\
F_0\left(k\right)&=&1,
\label{eq:expansion_of_self_enargy_0}
\end{eqnarray}
where for $\sigma\in\theta$,
 $\theta_{+}\equiv\theta\cup\left\{\left\{0\right\}\right\}$,
\begin{eqnarray}
  \tilde{S}\left(\left[\lambda,\theta\right],k,\sigma\right) &\equiv&
  \int\prod_{\sigma'\in
    \theta,\sigma'\neq\sigma}\frac{dk_{\sigma'}}{2\pi}
  \left(
    \prod_{\left\{\sigma',\sigma''\right\}\in\lambda}
    K\left(k_{\sigma'-k_\sigma''}\right)
  \right)
\nonumber\\&&{}\times
  \exp
  \left(
    -\beta\sum_{\sigma'\in\theta,\sigma'\neq\sigma}
    \#\sigma'
    k^2_{\sigma'}
  \right).
\end{eqnarray}

It can be easily checked that (\ref{eq:expansion_of_self_enargy_0}) is true.

From eq.(\ref{eq:self_enargy}), a recursive relation for
$F_N\left(k\right)$ is obtained as
\begin{eqnarray}
F_N\left(k\right)
&=&
\frac{1}{N!}
\sum_{\theta\in\Theta\left(N\right)}
\prod_{\sigma\in\theta}
\int\frac{dq_\sigma}{2\pi}K\left(k-q_\sigma\right)
\nonumber\\&&{}
\sum_{\theta^{\prime}\in\Theta\left(\#\sigma\right)}
\left(\#\theta^{\prime}-1\right)!
\exp\left(-\beta \#\theta^{\prime} q^2_\sigma\right)
\prod_{\sigma^{\prime}\in\theta^{\prime}}
\#\sigma^{\prime}! F_{\#\sigma^{\prime}-1}\left(q_\sigma\right).
\label{eq:F_recusive_formula}
\end{eqnarray}
Substituting the expression of $F_N\left(k\right)$ in
(\ref{eq:expansion_of_self_enargy}) into $F_{\#\sigma-1}\left(q\right)$ in
(\ref{eq:F_recusive_formula}), we obtain
\begin{eqnarray}
F_N\left(k\right)
&=&
\frac{1}{N!}
\sum_{\theta\in\Theta\left(N\right)}
\prod_{\sigma\in\theta}
\int\frac{dq_\sigma}{2\pi}K\left(k-q_\sigma\right)
\sum_{\theta^{\prime}\in\Theta\left(\#\sigma\right)}
\left(\#\theta^{\prime}-1\right)!
\exp\left(-\beta \#\theta^{\prime} q^2_\sigma\right)
\nonumber\\&&{}
\prod_{\sigma^{\prime}\in\theta^{\prime}}
\#\sigma^{\prime}
\sum_{\theta^{\prime\prime}\in\Theta\left(\#\sigma^{\prime}-1\right)}
\sum_
{
  \lambda''\in\Lambda_c\left(\theta^{\prime\prime}_{+}\right)
}
\nonumber\\&&{}
\left(
  \prod_{\sigma\in\theta^{\prime\prime},\sigma \neq \left\{0\right\}}
  \left(\#\sigma-1\right)!
  \#\sigma^{n\left(\sigma,\lambda^{\prime\prime}\right)-1}
\right)
\tilde{S}\left(\left[\lambda'',\theta''_+\right],q_\sigma,\left\{0\right\}\right).
\label{eq:F_recusive_formula_e-}
\end{eqnarray}
For non-integer argument of $\Theta\left(\sigma\right)$,
 (\ref{eq:F_recusive_formula_e-}) becomes
\begin{eqnarray}
F_N\left(k\right)
&=&
\frac{1}{N!}
\sum_{\theta\in\Theta\left(N\right)}
\prod_{\sigma\in\theta}
\int\frac{dq_\sigma}{2\pi}K\left(k-q_\sigma\right)
\sum_{\theta^{\prime}\in\Theta\left(\sigma\right)}
\left(\#\theta^{\prime}-1\right)!
\exp\left(-\beta \#\theta^{\prime} q^2_\sigma\right)
\nonumber\\&&{}
\prod_{\sigma^{\prime}\in\theta^{\prime}}
\sum_{n\in\sigma}
\sum_{\theta''\in\Theta\left(\sigma^{\prime}-\left\{n\right\}\right)}
\sum_
{
  \lambda''\in\Lambda_c\left(\theta^{\prime\prime}\cup\left\{\left\{n\right\}\right\}\right)
}
 \nonumber\\&&{}
\left(
  \prod_{\sigma\in\theta^{\prime\prime},\sigma \neq \left\{0\right\}}
  \left(\#\sigma-1\right)!
  \#\sigma^{n\left(\sigma,\lambda^{\prime\prime}\right)-1}
\right)
\tilde{S}\left(\left[\lambda'',\theta''\cup\left\{\left\{n\right\}\right\}\right],q_\sigma,\left\{n\right\}\right).
\label{eq:F_recusive_formula_e}
\end{eqnarray}

We can show that
\begin{eqnarray}
&&
\sum_{\theta''\in\Theta\left(N\right)}
f_1\left(\#\theta\right)
\prod_{\sigma''\in\theta''}
\sum_{n\in\sigma''}
\sum_{\theta'''\in\Theta\left(\sigma''-\left\{n\right\}\right)}
\sum_
{
  \lambda'''\in\Lambda_c\left(\theta'''\cup\left\{\left\{n\right\}\right\}\right)
}
\nonumber\\&&
f_2\left(\left[\lambda''',\theta'''\cup\left\{\left\{n\right\}\right\}\right],\left\{n\right\}\right)
\prod_{\sigma'\in\theta''',\sigma'\neq\left\{n\right\}}
f_3\left(n\left(\sigma',\lambda'''\right),\sigma'\right) \nonumber\\
&=&
\sum_{\theta\in\Theta\left(N\right)}
\sum_
{
  \lambda\in\Lambda_c\left(\theta\right)
}
\sum_{\sigma\in\theta}
\#\sigma^{n\left(\sigma,\lambda\right)}
f_1\left(\#\sigma\right)
f_2\left(\left[\lambda,\theta\right],\sigma\right)
\prod_{\sigma'\in\theta,\sigma'\neq\sigma}
f\left(n\left(\sigma',\lambda\right),\sigma'\right)
  \label{eq:combinatorial_d4}
\end{eqnarray}
where $f_1\left(n\right)$ and $f_3\left(n,\sigma\right)$ are arbitrary
functions, and $f_2\left(\left[\lambda,\theta\right]\right)$ is a
function which satisfies
\begin{eqnarray}
  f_2\left(\left[\lambda,\theta\right],\sigma\right)
  &=&
  \prod_{i}f_2\left(\left[\lambda_i,\theta_i\right],\sigma_i\right)
\end{eqnarray}
\begin{eqnarray}
  \lambda
  =
  \bigcup_i  \lambda_i
  \;,\quad
  \theta
  =
  \bigcup_i\left(\theta_i-\left\{\sigma_i\right\}\right)\quad
  \cup\quad\Bigr\{\bigcup_i \sigma_i\Bigr\}.\nonumber
\end{eqnarray}
The meaning of the right hand side of (\ref{eq:combinatorial_d4}) is
to sum up all the patterns generated by the following
process:
first, divide a set $\left\{1,\cdots,N\right\}$ into sets of $\theta$,
second define a connection pattern $\lambda$ which consists of one cluster,
then choose one set $\sigma$ in $\theta$.
The meaning of the left hand side of (\ref{eq:combinatorial_d4}) is to sum up all the patterns 
generated by the following process:
first, divide $\left\{1,\cdots,N\right\}$ into sets of $\theta''$ each
of which contain one element $n$ in the $\sigma$ chosen in
r.h.s. and elements of which some sets in $\theta$ consists.  The
$\sigma$ chosen in r.h.s. is not one of these sets, and these sets are
connected with each other by connections in $\lambda$ and any sets are
connected with the $\sigma$ by connections in $\lambda$.
Second, divide each of sets in $\theta''$ into sets of $\theta'$ and
$\left\{n\right\}$.
Then, define a connection pattern $\lambda''$ of sets in $\theta'''$
and $\left\{n\right\}$. The connection pattern $\lambda''$ consists of
one cluster.
Fig.\ref{fig:combi_3l} and Fig.\ref{fig:combi_3r} illustrate graphical
representations of both sides of patterns in
(\ref{eq:combinatorial_c3}).  Note that if one chooses a
r.h.s. pattern, there exist the corresponding $\#\sigma^{n\left(\sigma,\lambda\right)}$
l.h.s. patterns.
\begin{figure}[p]
\begin{center}
\begin{psfrags}
      \psfrag{aa}{$\theta''$}
      \psfrag{aba}{$\theta'''_1$}
      \psfrag{abb}{$\theta'''_2$}
      \psfrag{abc}{$n_1$}
      \psfrag{abd}{$n_2$}
      \psfrag{aca}{$\left[\lambda'''_1,\theta'''_1\cup\left\{\left\{n_1\right\}\right\}\right]$}
      \psfrag{acb}{$\left[\lambda'''_2,\theta'''_2\cup\left\{\left\{n_2\right\}\right\}\right]$}
      \psfrag{ba}{$\Theta\left(\sigma''-\left\{n\right\}\right)$}
      \psfrag{bb}{$\Lambda_c\left(\theta'''\cup\left\{\left\{n\right\}\right\}\right)$}
  \psfig{file=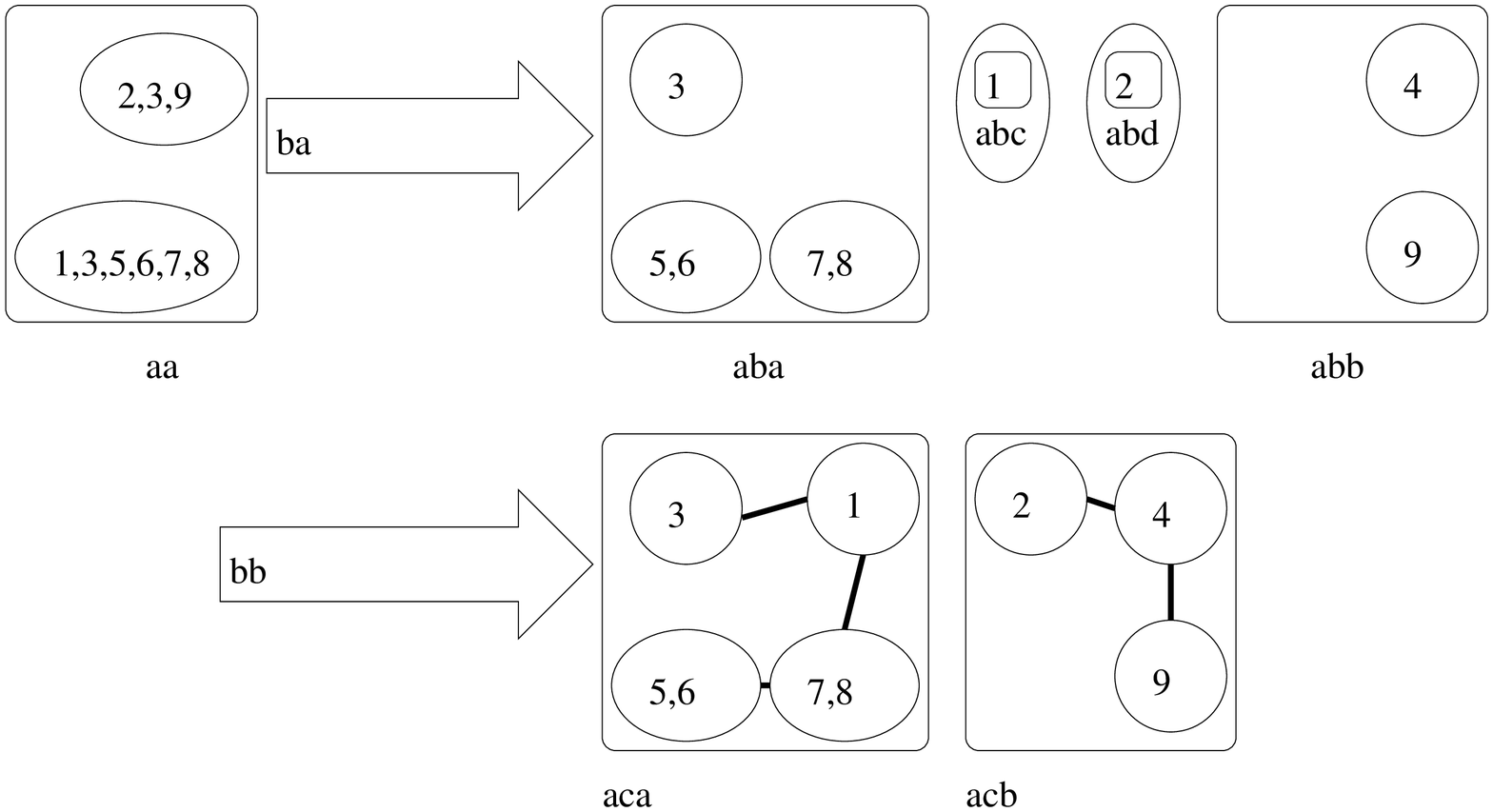 , scale = 0.50}
\end{psfrags}
\caption{A graphical representation of one of  patterns which are summed up in the left hand side of eq.(\ref{eq:combinatorial_d4}).}
\label{fig:combi_4l}
\begin{psfrags}
      \psfrag{aa}{$\theta$}
      \psfrag{ab}{$\left[\lambda,\theta\right]$}
      \psfrag{ac}{$\sigma$}
      \psfrag{ba}{$\Lambda_c\left(\theta\right)$}
      \psfrag{bb}{$\theta$}
  \psfig{file=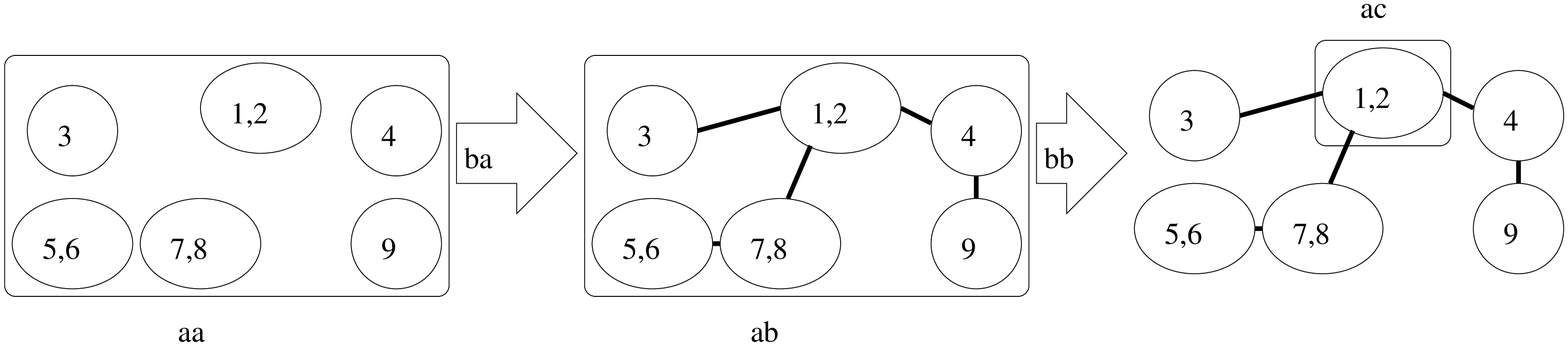 , scale = 0.50}
\end{psfrags}
\caption{A graphical representation of one of patterns which are summed up in the right hand side of eq.(\ref{eq:combinatorial_d4}).}
\label{fig:combi_4r}
\end{center}
\end{figure}

Applying the relation (\ref{eq:combinatorial_d4}) to (\ref{eq:F_recusive_formula_e}), we have
\begin{eqnarray}
  \label{eq:F_recusive_formula_e_2}
  F_N\left(k\right)
  &=&
\frac{1}{N!}
\sum_{\theta\in\Theta\left(N\right)}
\prod_{\sigma\in\theta}
\int\frac{dq_\sigma}{2\pi}K\left(k-q_\sigma\right)
\nonumber\\&&{}
\sum_{\theta^{\prime}\in\Theta\left(\sigma\right)}
\sum_{\lambda'\in\Lambda_c\left(\theta'\right)}
\sum_{\sigma'\in\theta'}
\#\sigma^{\prime n\left(\sigma',\lambda'\right)}
\left(\#\sigma'-1\right)!
\exp\left(-\beta \#\sigma' q^2_\sigma\right)
\tilde{S}\left(\left[\lambda',\theta'\right],q_\sigma,\sigma'\right)
\nonumber\\&&{}
\prod_{\sigma''\in\theta',\sigma''\neq\sigma'}
  \left(\#\sigma''-1\right)!
  \#\sigma^{\prime\prime n\left(\sigma'',\lambda'\right)-1}.
\nonumber\\
   &=&
\frac{1}{N!}
\sum_{\theta\in\Theta\left(N\right)}
\prod_{\sigma\in\theta}
\sum_{\theta^{\prime}\in\Theta\left(\sigma\right)}
\sum_{\lambda'\in\Lambda_c\left(\theta'\right)}
\sum_{\sigma'\in\theta'}
\nonumber\\&&{}
\int\frac{dq_{\sigma'}}{2\pi}K\left(k-q_{\sigma'}\right)
\#\sigma^{\prime n\left(\sigma',\lambda'\right)}
\left(\#\sigma'-1\right)!
\exp\left(-\beta \#\sigma' q^2_{\sigma'}\right)
\tilde{S}\left(\left[\lambda',\theta'\right],q_{\sigma'},\sigma'\right)
\nonumber\\&&{}
\prod_{\sigma''\in\theta',\sigma''\neq\sigma'}
  \left(\#\sigma''-1\right)!
  \#\sigma^{\prime\prime n\left(\sigma'',\lambda'\right)-1}.
\nonumber\\
  &=&
\frac{1}{N!}
\sum_{\theta\in\Theta\left(N\right)}
\sum_{\lambda\in\Lambda\left(\theta\right)}
\prod_{\left[\lambda',\theta'\right]\in\tilde{G}\left(\left[\lambda,\theta\right]\right)}
\sum_{\sigma'\in\theta'}
\nonumber\\&&{}
\int\frac{dq_{\sigma'}}{2\pi}K\left(k-q_{\sigma'}\right)
\#\sigma^{\prime n\left(\sigma',\lambda'\right)}
\left(\#\sigma'-1\right)!
\exp\left(-\beta \#\sigma' q^2_{\sigma'}\right)
\tilde{S}\left(\left[\lambda',\theta'\right],q_{\sigma'},\sigma'\right)
\nonumber\\&&{}
\prod_{\sigma''\in\theta',\sigma''\neq\sigma'}
  \left(\#\sigma''-1\right)!
  \#\sigma^{\prime\prime n\left(\sigma'',\lambda'\right)-1}
\label{eq:tmp_proof_d_1}
\end{eqnarray}
The second equality is due to a replacement of integration variables
from $q_\sigma$ to $q_{\sigma'}$, and the relation
(\ref{eq:combinatorial_c1}) is used in the third equality.

By definitions, the following relation can be shown,
\begin{eqnarray}
\!\!\!\!\!\!\!\!
\!\!\!\!\!\!\!\!
\!\!\!\!\!\!\!\!
\tilde{S}
\left(
  \left[\lambda,\theta\right],
  k,
  \left\{0\right\}
\right)
&=&
\prod_{\left[\lambda_i,\theta_i\right]\in\tilde{G}\left(\left[\lambda,\theta\right]\right)}
\int\frac{dq_{\sigma_i}}{2\pi}K\left(k-q_{\sigma_i}\right)
\exp\left(-\beta \#\sigma' q^2_{\sigma_i}\right)
\tilde{S}\left(\left[\lambda_i,\theta_i\right],q_{\sigma_i},\sigma_i\right),
\label{eq:tmp_proof_d_2}
\end{eqnarray}
\begin{eqnarray}
\lambda
=
\bigcup_{i}\lambda_i\;
\cup\;
\bigcup_i\left\{\xi_i\right\},\quad
\xi_i=\left\{\sigma_i,\left\{0\right\}\right\},\quad
\theta
=
\bigcup_i \theta_i \;
\cup\;
\left\{\left\{0\right\}\right\}.\nonumber
\end{eqnarray}

Using a relation (\ref{eq:tmp_proof_d_2}) in (\ref{eq:tmp_proof_d_1}), we obtain
\begin{eqnarray}
\!\!\!\!\!\!\!\!
\!\!\!\!\!\!\!\!
  F_N\left(k\right)
&=&
\frac{1}{N!}
\sum_{\theta\in\Theta\left(N\right)}
\sum_
{
  \lambda\in\Lambda_c\left(\theta_{+}\right)
}
\left(
  \prod_{\sigma\in\theta,\sigma \neq \left\{0\right\}}
  \left(\#\sigma-1\right)!
  \#\sigma^{n\left(\sigma,\lambda\right)-1}
\right)
\tilde{S}\left(\left[\lambda,\theta\right],k,\left\{0\right\}\right).
  \label{eq:F_recusive_formula_e_2}
\end{eqnarray}
We see that (\ref{eq:F_recusive_formula_e_2}) is equal to (\ref{eq:expansion_of_self_enargy}). Therefore, (\ref{eq:expansion_of_self_enargy}) is recursively proved. 

From (\ref{eq:integrant_of_grand_partition_function}), we have
\begin{eqnarray}
b_N
&=&
\int\frac{dk}{2\pi}
\sum_{\theta\in\Theta\left(N\right)}
\exp\left(-\#\theta\beta k^2\right)
\frac{\left(\#\theta - 1\right)! }{N!}
\prod_{\sigma\in\theta}
\#\sigma!F_{\#\sigma-1}\left(k\right).
\label{eq:b_recusive_formula}
\end{eqnarray}
Substituting the expression of $F_N$ in
(\ref{eq:expansion_of_self_enargy}) into $F_{\#\sigma-1}$ in
(\ref{eq:b_recusive_formula}) gives
\begin{eqnarray}
b_N
&=&
\int\frac{dk}{2\pi}
\sum_{\theta\in\Theta\left(N\right)}
\exp\left(-\#\theta\beta k^2\right)
\frac{\left(\#\theta - 1\right)! }{N!}
\prod_{\sigma\in\theta}
\#\sigma
\sum_{\theta^{\prime}\in\Theta\left(\#\sigma-1\right)}
\sum_{\lambda^{\prime}\in\Lambda,\left(\theta^{\prime}_{+}\right)
}
\nonumber\\&&{}
\left(
  \prod_{\sigma^{\prime}\in\theta^{\prime},\sigma^{\prime} \neq \left\{0\right\}}
  \left(\#\sigma^{\prime}-1\right)!
  \#\sigma^{n\left(\sigma^{\prime},\lambda^{\prime}\right)-1}
\right)
\tilde{S}\left(\left[\lambda^{\prime},\theta^{\prime}\right],k,\left\{0\right\}\right).
\label{eq:tmp_2}
\end{eqnarray}
Using the relation (\ref{eq:combinatorial_d4}), we have
\begin{eqnarray}
b_N
&=&
\sum_{\theta\in\Theta\left(N\right)}
\sum_{\lambda\in\Lambda_c\left(\theta\right)}
\sum_{\sigma\in\theta}
\int\frac{dk}{2\pi}
\exp\left(-\#\sigma\beta k^2\right)
\#\sigma^{\prime n \left(\sigma,\lambda\right)}
\frac{\left(\#\sigma - 1\right)! }{N!}
\tilde{S}\left(\left[\lambda,\theta\right],k,\sigma\right)
\nonumber\\&&{}
\prod_{\sigma'\in\theta,\sigma' \neq \sigma}
  \left(\#\sigma'-1\right)!
  \#\sigma^{\prime n \left(\sigma^{\prime},\lambda\right)-1}.
\label{eq:tmp_proof_d_3}
\end{eqnarray}
Since the following relation holds,
\begin{eqnarray}
S
\left(
  \left[\lambda,\theta\right]
\right)
&=&
\int\frac{dq_{\sigma}}{2\pi}K\left(k-q_{\sigma}\right)
\exp\left(-\beta \#\sigma' q^2_{\sigma}\right)
\tilde{S}\left(\left[\lambda,\theta\right],q_{\sigma},\sigma\right),
\label{eq:tmp_proof_d_4}
\\&&
\lambda
\in\Lambda_c\left(\theta\right),\quad
\sigma\in\theta
,\nonumber
\end{eqnarray}
we arrive at
\begin{eqnarray}
b_N
&=&
\frac{1}{N!}
\sum_{\theta\in\Theta\left(N\right)}
\sum_{\lambda\in\Lambda_c\left(\theta\right)}
\left(\sum_{\sigma\in\theta}\#\sigma\right)
\left(
  \prod_{\sigma\in\theta}
  \left(\#\sigma-1\right)!
  \#\sigma^{n\left(\sigma,\lambda\right)-1}
\right)
S^{\prime}_{\infty}\left(\left[\lambda,\theta\right]\right)\\
&=&
\frac{1}{\left(N-1\right)!}
\sum_{\theta\in\Theta\left(N\right)}
\sum_{\lambda\in\Lambda\left(\theta\right)}
\left(
  \prod_{\sigma\in\theta}
  \left(\#\sigma-1\right)!
  \#\sigma^{n\left(\sigma,\lambda\right)-1}
\right)
S^{\prime}_{\infty}\left(\left[\lambda,\theta\right]\right),
\end{eqnarray}
where $S^{\prime}_{\infty}\left(\lambda,\theta\right)$ is defined in
(\ref{eq:define_S_prime_inf}).  This expression of the cluster integral $b_N$ is the same as
(\ref{eq:cluster_integral_22}).

\end{document}